\documentclass[11pt,a4paper]{scrartcl}
\usepackage[round,authoryear]{natbib}

\usepackage[top=3cm, bottom=3cm, left=3cm, right=3cm]{geometry}

\usepackage{booktabs}

\usepackage[table]{xcolor}
\usepackage{color}
\usepackage{latexsym}              
\usepackage{amsmath}               
\usepackage{amssymb}               
\usepackage{amsfonts}              
\usepackage{amsthm}                
\usepackage{dsfont}
\usepackage{multirow}
\usepackage{tikz}                  
\usetikzlibrary{arrows,positioning,shapes,bayesnet}
\usepackage{tcolorbox}

\usepackage{xspace} 
\usepackage{float}
\usepackage{bbold}
\usepackage{cuted} 
\usepackage{algorithm, algorithmic}
\restylefloat{table}

\usepackage[english]{babel} 

\RequirePackage[%
  pdfstartview=FitH,%
  breaklinks=true,%
  bookmarks=true,%
  colorlinks=true,%
  linkcolor= blue,
  anchorcolor=blue,%
  citecolor=blue,
  filecolor=blue,%
  menucolor=blue,%
  urlcolor=blue%
  ]{hyperref}
  
  \AtBeginDocument{%
  \hypersetup{%
    pdfauthor={},%
    	colorlinks = true,%
  	urlcolor = blue,%
  	linkcolor = blue,%
  	citecolor = orange,%
    pdftitle={Compilation: \today}%
  }
}

\usepackage[mathcal]{}
\usepackage{cleveref}
\crefname{assumption}{Assumption}{Assumptions}
\crefname{equation}{Eq.}{Eqs.}
\crefname{figure}{Fig.}{Figs.}
\crefname{table}{Table}{Tables}
\crefname{section}{Sec.}{Secs.}
\crefname{theorem}{Thm.}{Thms.}
\crefname{lemma}{Lemma}{Lemmas}
\crefname{corollary}{Cor.}{Cors.}
\crefname{example}{Example}{Examples}
\crefname{appendix}{Appendix}{Appendixes}
\crefname{remark}{Remark}{Remark}

\renewenvironment{proof}[1][\proofname]{{\bfseries #1.}}{\qed \\ }

\makeatother

\theoremstyle{plain}  
\newtheorem{theorem}{Theorem}[section]

\newtheorem{lemma}[theorem]{Lemma}
\newtheorem{proposition}[theorem]{Proposition}

\newcommand{\paren}[1]{\left( #1 \right)}
\newcommand{\croch}[1]{\left[\, #1 \,\right]}
\newcommand{\acc}[1]{\left\{ #1 \right\}}
\newcommand{\abs}[1]{\left| #1 \right|}

\newcommand{\argmax}{\operatornamewithlimits{argmax}} 
\newcommand{\dif}{\mathop{}\!\mathrm{d}}
\renewcommand\hat{\widehat}
\newcommand{\indep}{\perp \!\!\! \perp}

\renewcommand{\vert}{\mid}

\newcommand{\I}{\mathcal{I}}

\newcommand{\algo}{\textsc{Magma}\xspace}

\newcommand\yi{\mathbf{y}_i}
\newcommand\yst{y_{*}}
\newcommand\yii{\acc{\mathbf{y}_i}_i}
\newcommand\ytaui{\textbf{y}_i^{\boldsymbol \tau_i}}
\newcommand\ti{\mathbf{t}_i}
\newcommand\Ut{\mathbf{t}}
\newcommand\tpred{\mathbf{t}^{p}}
\newcommand\tst{\mathbf{t}_{*}}
\newcommand\tpst{\mathbf{t}^{p}_{*}}

\newcommand\mhat{\hat{m}_0}

\newcommand\mutau{\boldsymbol{\mu}_0^{\boldsymbol \tau}}
\newcommand\mutaui{\boldsymbol{\mu}_0^{\boldsymbol \tau_i}}

\newcommand\taub{\boldsymbol\tau}
\newcommand\mtau{\mathbf{m}_0^{\boldsymbol \tau}}
\newcommand\psii{\psi_{\theta_i, \sigma_i^2}}
\newcommand\psiihat{\psi_{\hat{\theta}_i, \hat{\sigma}_i^2}}
\newcommand\Psii{\boldsymbol{\Psi}_{\theta_i, \sigma_i^2}}

\newcommand\Psiihat{\boldsymbol{\Psi}_{\hat{\theta}_i, \hat{\sigma}_i^2}}

\newcommand\Kthetahat{\mathbf{K}_{\hat{\theta}_0}}
\newcommand\Khat{\hat{\mathbf{K}}}
\newcommand\thetaii{\acc{\theta_i}_i}

\newcommand\sigmaii{\acc{\sigma_i^2}_i}

\newcommand\sumi{\sum\limits_{i = 1}^{M}}

\newcommand\sumu{\sum\limits_{u \in \taub}}
\newcommand\sumv{\sum\limits_{v \in \taub}}

\begin{document}

\title{MAGMA: Inference and Prediction using Multi-Task Gaussian Processes with Common Mean}



\title{MAGMA: Inference and Prediction using Multi-Task Gaussian Processes with Common Mean}
\author{
\textbf{Arthur Leroy}\footnote{This work has been carried out while the author was affiliated with Université de Paris, CNRS, MAP5 UMR 8145, F-75006 Paris, France} \\ [2ex]
The University of Sheffield, Department of Computer Science, \\
Sheffield, United Kingdom, \\
arthur.leroy.pro@gmail.com \\\\
\textbf{Pierre Latouche} \\ [2ex]
Universit\'e de Paris, CNRS, MAP5 UMR 8145, \\
F-75006 Paris, France \\
pierre.latouche@math.cnrs.fr \\\\
\textbf{Benjamin Guedj} \\ [2ex]
Inria, France and \\
University College London, United Kingdom \\
benjamin.guedj@inria.fr \\\\
\textbf{Servane Gey} \\ [2ex]
Universit\'e de Paris, CNRS, MAP5 UMR 8145, \\
F-75006 Paris, France \\
servane.gey@parisdescartes.fr
}
\date{\today}

\maketitle

\begin{abstract}
A novel multi-task Gaussian process (GP) framework is proposed, by using a common mean process for sharing information across tasks.
In particular, we investigate the problem of time series forecasting, with the objective to improve multiple-step-ahead predictions. 
The common mean process is defined as a GP for which the hyper-posterior distribution is tractable.
Therefore an EM algorithm is derived for handling both hyper-parameters optimisation and hyper-posterior computation. 
Unlike previous approaches in the literature, the model fully accounts for uncertainty and can handle irregular grids of observations while maintaining explicit formulations, by modelling the mean process in a unified GP framework.
Predictive analytical equations are provided, integrating information shared across tasks through a relevant prior mean.
This approach greatly improves the predictive performances, even far from observations, and may reduce significantly the computational complexity compared to traditional multi-task GP models.
Our overall algorithm is called \textsc{Magma} (standing for Multi tAsk Gaussian processes with common MeAn). 
The quality of the mean process estimation, predictive performances, and comparisons to alternatives are assessed in various simulated scenarios and on real datasets.
\end{abstract}


\section{Introduction}
\label{sec:intro}

Gaussian processes (GPs) are a powerful tool, widely used in machine learning \citep{BishopPatternrecognitionmachine2006, RasmussenGaussianprocessesmachine2006a}. 
The classic context of regression aims at inferring the underlying mapping function associating input to output data.
In a probabilistic framework, a typical strategy is to assume that this function is drawn from a prior GP.
Doing so, we may enforce some properties for the function solely by characterising the mean and covariance functions of the process, the latter often being associated with a specific kernel. 
This covariance function plays a central role and GPs are an example of kernel methods.
We refer to \cite{AlvarezKernelsVectorValuedFunctions2012a} for a comprehensive review.
On the other hand, the mean function is generally set to 0 for all entries assuming that the covariance structure already integrates the desired relationship between observed data and prediction targets. 
In this paper, we consider a novel multi-task learning framework where a series of GPs share a common mean, expressed as a GP as well.
We demonstrate that modelling the mean function as such can be key to obtain more relevant predictions.

\paragraph{Related work} 
~\par

The multi-task framework consists in using data from several tasks (or individuals) to improve learning or predictive capacities compared to an isolated model. 
It has been introduced by \cite{CaruanaMultitaskLearning1997} and then adapted in many fields of machine learning. 
GP versions of such models were introduced in \cite{SchwaighoferLearningGaussianProcess2004}, which proposed an Expectation-Maximisation (EM) algorithm for learning. 
Similar techniques can be found in \cite{ShiHierarchicalGaussianprocess2005}.
Meanwhile, \cite{YuLearningGaussianProcesses2005} offered an extensive study of the relationships between the linear model and GPs to develop a multi-task GP formulation. 
However, since the introduction in \cite{BonillaMultitaskGaussianProcess2008} of the idea of two matrices, modelling covariance between inputs and tasks respectively, the term \emph{multi-task Gaussian process} has mostly referred to the choice made regarding the covariance structure.  
Some further developments were discussed by \cite{HayashiSelfmeasuringSimilarityMultitask2012}, \cite{RakitschItallnoise2013} and \cite{ZhuMultitaskSparseGaussian2014}.
In particular, an interesting approach in \cite{NguyenCollaborativemultioutputGaussian2014} proposed a sparse approximation for multi-task GP inference.
More generally, these approaches are known as examples of \emph{linear models of coregionalization} (LMC) in the geostatistics literature, and \cite{AlvarezComputationallyEfficientConvolved2011a} provides a unified view on the topic as well as an efficient strategy for constructing computationally efficient approximations.
Let us emphasise that the present paper is not based on the same assumptions and principles, and aims at defining a different multi-task paradigm for GPs, focusing on sharing information through the mean function rather than the covariance structure. 
Besides, the work of \cite{SwerskyMultiTaskBayesianOptimization2013} on Bayesian hyper-parameter optimisation in such LMC models is also worth a mention.
Real applications were tackled by similar models in \cite{WilliamsMultitaskGaussianProcess2009} and \cite{AlaaBayesianInferenceIndividualized2017}, while \cite{ClingermanLifelongLearningGaussian2017} and \cite{Moreno-MunozContinualMultitaskGaussian2019a} developed continual learning methods for multi-task GP.
\newline

As we focus on multi-task time series forecasting, a connection can be drawn to the study of multiple curves, or functional data analysis (FDA). 
As initially proposed in \cite{RiceEstimatingMeanCovariance1991}, it is possible to model and learn mean and covariance structures simultaneously in this context. 
We refer to the monographs \cite{RamsayFunctionalDataAnalysis2005} and \cite{FerratyNonparametricFunctionalData2006} for a comprehensive introduction to FDA.
In particular, these books introduced several usual ways for modelling a set of functional objects in frequentist frameworks, for example by using a decomposition in a basis of functions (such as B-splines, wavelets, Fourier). 
This kind of B-splines decomposition was used in \cite{ShiGaussianProcessFunctional2007} for modelling the mean function in a generative model that somehow resembles ours.
Subsequently, some Bayesian alternatives were developed in \cite{ThompsonBayesianModelSparse2008}, and \cite{CrainiceanuBayesianFunctionalData2010}.

\paragraph{Our contributions}
~\par

A multi-task GP framework with a common mean process is introduced, allowing reliable probabilistic forecasts even in multiple-step-ahead problems, or for sparsely observed individuals. 
For this purpose,
(i) we introduce a GP model where the specific covariance structure of each task is defined through a separate kernel and its associated set of hyper-parameters, whereas the common mean function $\mu_0$ allows sharing information across tasks and overcomes the weaknesses of classic GPs in making predictions far from observed data.
To account for uncertainty, we propose a hierarchical formulation to define the common mean process $\mu_0$ as a GP as well.
(ii) We derive an algorithm called \algo (available as an R package at \url{https://github.com/ArthurLeroy/MagmaClustR}) to compute $\mu_0$'s hyper-posterior distribution together with the estimation of hyper-parameters in an EM fashion, and discuss its computational complexity.
(iii) We enrich \algo with explicit formulas to make predictions for any new, partially observed, task.
The hyper-posterior distribution of $\mu_0$ provides a prior belief on what we would expect to observe before seeing any new data, acting as an already-informed mean process, integrating both trend and uncertainty coming from other tasks.
(iv) We illustrate the performance of our method on synthetic and two real-life datasets and obtain state-of-the-art results compared to alternative approaches.

\paragraph{Outline} 
~\par

The paper is organised as follows.
We introduce our multi-task Gaussian process model in \Cref{sec:model}, along with notation. \Cref{sec:inference} is devoted to the inference procedure, with an Expectation-Maximisation (EM) algorithm to estimate the Gaussian process hyper-parameters and $\mu_0$'s hyper-posterior.
We leverage this strategy in \Cref{sec:prediction} and derive a prediction algorithm. 
In \Cref{sec:complexity}, we analyse and discuss the computational complexity of both the inference and prediction procedures. 
Our methodology is illustrated in \Cref{sec:exp}, with a series of experiments on both synthetic and real-life datasets, and a comparison to competing state-of-the-art algorithms. On those tasks, we provide empirical evidence that our algorithm outperforms other approaches. \Cref{sec:conclusion} draws perspectives for future work, and we defer some proofs to original results claimed in the paper to \Cref{sec:proofs}.

\section{The model}
\label{sec:model}
\subsection{Notation}
\label{sec:notation}

While GPs can handle many types of data, their continuous nature makes them particularly well suited to study temporal phenomena.
Throughout, the term \emph{individual} is used as a synonym of \emph{task} or \emph{batch}, and we adopt notation and vocabulary of time series to remain consistent with the application on real dataset provided in \Cref{sec:simu_real_data}, which addresses young swimmers performances' forecast.
\newline

We are provided with functional data coming  from  $M  \in
\mathcal{I}$ different individuals, where $\mathcal{I} \subset \mathbb{N}$.  
For each individual $i$, we observe a set of inputs $\{ t_i^1, \dots , \allowbreak t_i^{N_i} \}$ and associated outputs $\{ y_i(t_i^1), \dots , y_i(t_i^{N_i}) \}$, where $N_i$ is the number of data points for the $i$-th individual.
Since many  objects are defined for all  individuals, we
shorten our notation as follows: for any  object $x$ existing
for all $i$, we denote $\acc{x_i}_i = \acc{x_1, \dots, x_M}$.
Moreover, as we work in a temporal context, the inputs are referred to as \textit{timestamps}.
In the specific case where all individuals are observed at the same timestamps, we call the grid of observations \textit{common}.
On the contrary, a grid of observations is \textit{uncommon} if the timestamps are different in number and/or location among the individuals.
Some convenient notation follows:

\begin{itemize}
    \item $\ti = \{ t_i^1,\dots,t_i^{N_i} \}$, the set of timestamps for the  $i$-th individual,
    \item $\yi  = y_i(\ti)$, the  vector of outputs for the $i$-th individual,
    \item $\Ut = \bigcup\limits_{i = 1}^M \ti$, \ the pooled set of timestamps among individuals,
    \item $N = \operatorname{card}(\Ut)$, the total number of observed timestamps.
\end{itemize}

\subsection{Model and hypotheses}
\label{sec:model_hypo}

\begin{figure}
\begin{center}
\begin{tikzpicture}
  
  \node[obs]                               (y) {$y_i$};
  
  \node[latent, above=of y, xshift=-1.35cm](mu){$\mu_0$};
  \node[const, above=of mu, xshift=-0.5cm] (m) {$m_0$};
  \node[const, above=of mu, xshift= 0.5cm] (t0) {$\theta_0$};

  \node[latent, above=of y, xshift=1.35cm] (f) {$f_i$};
  \node[const, above=of f, xshift= 0cm]	   (ti) {$\theta_i$};
    
  \node[latent, right= 1cm of y]           (e) {$\epsilon_i$};
  \node[const, right= 1cm of e]		   (s) {$\sigma_i^2$};

  \factor[above=of mu] {mu-f} {left:$\mathcal{N}$} {m,t0} {mu} ;
  \factor[above=of f] {f-f} {left:$\mathcal{N}$} {ti} {f} ;
  \factor[right=of e] {e-f} {above:$\mathcal{N}$} {s} {e} ;

  \edge {f,mu,e} {y} ;

  \plate {} {(f)(y)(e)(ti)(s)} {$\forall i \in \mathcal{I}$} ;

\end{tikzpicture}
\caption{Graphical model of dependencies between variables in the Multi-task Gaussian Process model.}
\label{graph_model} 
\end{center}     
\end{figure}
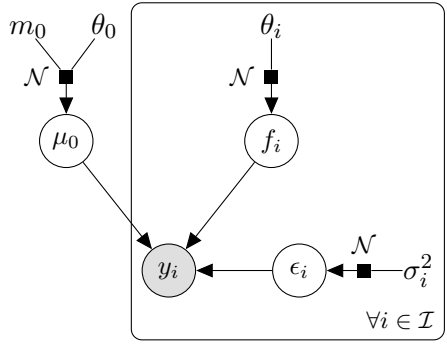

Suppose that functional data are coming from the sum of a mean process, common to all individuals, and an individual-specific centred process.
To clarify relationships in the generative model, we illustrate our graphical model in \Cref{graph_model}.
Let $\mathcal{T}$ be the input space, our model is
\begin{equation*}
    y_i(t) = \mu_0(t) + f_i(t) + \epsilon_i(t), \ \ \forall t \in \mathcal{T}, \ \ \forall i \in \I, 
\end{equation*}
where $\mu_0(\cdot) \sim \mathcal{GP} (m_0(\cdot), k_{\theta_0}(\cdot,\cdot))$ and $f_i(\cdot) \sim  \allowbreak \mathcal{GP} \paren{0,c_{\theta_i}(\cdot,\cdot)}$ are respectively the common mean and individual specific processes.
Moreover, the error term is supposed to be $\epsilon_i(\cdot) \sim \mathcal{N} (0,\sigma_i^2 I)$.
The following notation is used for parameters:


\begin{itemize}
	\item $m_0(\cdot)$, an arbitrary prior mean function,
 	\item $k_{\theta_0}(\cdot, \cdot)$, a covariance kernel of hyper-parameters $\theta_0$,
    \item $\forall i \in \I, \ c_{\theta_i}(\cdot, \cdot)$, a covariance kernel with hyper-parameters $\theta_i$,
    \item $\sigma_i^2 \in \mathbb{R}^{+}$, the noise variance associated with the $i$-th individual,
    \item $\forall i  \in \I,$ we define the shorthand $\psii(\cdot,\cdot) = c_{\theta_i}(\cdot,\cdot) + \sigma_i^2 I$,
    \item $\Theta = \{\theta_0, \acc{\theta_i}_i, \acc{\sigma_i^2}_i \}$, the set of all hyper-parameters to learn in the model.
\end{itemize}

\noindent We also assume that:
\begin{itemize}
    \item $\{ f_i \}_{i}$ are independent,
    \item $\{ \epsilon_i \}_{i}$ are independent,
    \item $\forall i \in \I, \ \mu_0$, $f_i$ and $\epsilon_i$ are independent.
\end{itemize}

\noindent It follows that $\{ y_i \vert \mu_0 \}_{i = 1,\dots,M}$ are independent from one another, and for all $i \in \mathcal{I}$: 
\begin{equation*}
    y_i(\cdot)  \vert \mu_0(\cdot) \sim \mathcal{GP}(\mu_0(\cdot), \psii(\cdot, \cdot)).
\end{equation*}

Let us emphasise that this property only holds conditionally to $\mu_0$. 
Otherwise, once $\mu_0$ is integrated out, the $y_i$ are no longer independent. 
Here, we do not assume any specific covariance structure between individuals contrarily to standard LMC approaches.
As we shall see in the next sections, the process $\mu_0$ will be key to handle the dependencies and share information across the individuals. 
\newline

Although this model is based on infinite-dimensional GPs, the inference will be conducted on a finite grid of observations.
According to the aforementioned notation, we observe $\{ (\ti, \yi) \}_{i}$, and the corresponding likelihoods are Gaussian: 
\begin{equation*}
    \yi \vert \mu_0(\ti) \sim \mathcal{N}(\yi; \mu_0(\ti), \Psii^{\ti}),
\end{equation*}
\noindent where  $\ \Psii^{\ti} = \psii (\ti, \ti) = \left[ \psii(k, l) \right]_{k, \ell \in \ti}$ is a $N_i \times N_i$ covariance matrix.
Since $\ti$ might be different among individuals, we also need to evaluate $\mu_0$ on the pooled grid of timestamps $\Ut$: 

\begin{equation*}
    \mu_0(\Ut) \sim \mathcal{N} \left(\mu_0(\Ut) ;  m_0(\Ut) , \mathbf{K}_{\theta_0}^{\Ut} \right),
\end{equation*}
\noindent where $\mathbf{K}_{\theta_0}^{\Ut} = k_{\theta_0}(\Ut, \Ut) = \left[ k_{\theta_0}(k, \ell) \right]_{k,l \in \Ut}$ is a $N \times N$ covariance matrix.   
\newline

An alternative hypothesis consists in considering hyper-parameters $\thetaii$ and $\sigmaii$ equal for all individuals.
We call this hypothesis \emph{Common HP} (where \emph{HP} stands for \emph{hyper-parameters}) in the \Cref{sec:exp}. 
This particular case represents a context where individuals correspond to different trajectories of the same process, whereas different hyper-parameters indicate different covariance structures and thus a more flexible model.
For the sake of generality, the remainder of the paper is written with $\theta_i$ and $\sigma_i^2$ notation, when there are no differences in the procedure. 
Moreover, the model above and the subsequent algorithm may use any form of covariance function, often parametrised by a finite set (usually small) of hyper-parameters. For example, a common kernel in the GP literature is known as the \emph{Exponentiated Quadratic} kernel (also called sometimes Squared Exponential or Radial Basis Function kernel). It solely depends on two hyper-parameters $\theta = \acc{v, \ell}$ and is defined as: 
\begin{equation}
\label{eq:kernel}
	k_{\mathrm{EQ}}\left(x, x^{\prime}\right)= v^{2} \exp \left(-\frac{\left(x-x^{\prime}\right)^{2}}{2 \ell^{2}}\right).
\end{equation}

The \emph{Exponentiated Quadratic} kernel is simple and enjoys useful smoothness properties. This is the kernel used in the current version of our implementation (see \Cref{sec:exp} for details). Note that there is a rich literature on kernel choice, their construction and properties, which is beyond the scope of the present work: we refer to \cite{RasmussenGaussianprocessesmachine2006a} or \cite{DuvenaudAutomaticmodelconstruction2014} for comprehensive studies.

\section{Inference}
\label{sec:inference}

\subsection{Learning}
\label{sec:EM_algo}

Several approaches to learn hyper-parameters for Gaussian processes have been proposed in the literature, we refer to \cite{RasmussenGaussianprocessesmachine2006a} for a comprehensive study.
One classical approach, called \emph{empirical Bayes} \citep{CasellaIntroductionEmpiricalBayes1985}, is based on the maximisation of an explicit likelihood to estimate hyper-parameters. 
This procedure avoids sampling from intractable distributions, usually resulting in additional computational cost and complicating practical use in moderate to large sample sizes. 
As previously stated, once $\mu_0$ is marginalised out, the log-likelihood cannot be written as a sum of Gaussian log-likelihoods any more.
Therefore, we propose an EM algorithm (see the pseudocode in \Cref{alg:algo_EM}) to learn the hyper-parameters $\Theta$ in this context. 
The procedure alternatively computes the hyper-posterior distribution $p(\mu_0 \vert (\yi)_i, \hat{\Theta})$ with current hyper-parameters, and then optimises $\Theta$ according to this hyper-posterior distribution. 
This EM algorithm converges to local maxima \citep{DempsterMaximumLikelihoodIncomplete1977}, typically in a handful of iterations. 

\paragraph{E step}
~\par

For the sake of simplicity, we assume in that section that $ \forall i,j \in \I, \ \ti = \mathbf{t}_j = \Ut$, i.e. the individuals are observed on a common grid of timestamps. We provide a generalisation of the following proposition in \Cref{sec:prediction} (\Cref{prop:post_mu}), where the result holds for uncommon grids.
The E step then consists in computing the hyper-posterior distribution of $\mu_0(\Ut)$.
\begin{proposition}
\label{prop:E_step}
Assume the hyper-parameters $\hat{\Theta}$ known from initialisation or estimated from a previous M step. The hyper-posterior distribution of $\mu_0$ remains Gaussian: 
\begin{equation}
p\paren{\mu_0(\Ut) \vert \yii, \hat{\Theta}} = \mathcal{N} \paren{\mu_0(\Ut) ; \mhat(\Ut), \Khat^{\Ut}},
\end{equation}
with
\begin{itemize}

\item $\Khat^{\Ut} = \paren{ { \Kthetahat^{\Ut}}^{-1} + \sumi {\Psiihat^{\Ut}}^{-1}}^{-1},$
\item $\mhat(\Ut) = \Khat^{\Ut} \paren{ { \Kthetahat^{\Ut}}^{-1} m_0\paren{\Ut} + \sumi { \Psiihat^{\Ut}}^{-1} \yi }.$
\end{itemize}

\end{proposition}
\begin{proof}

We omit specifying timestamps in what follows since each process is evaluated on $\Ut$. Therefore, we can write:
\begin{align*}
p \paren{ \mu_0 \vert \yii, \hat{\Theta} } 
& \propto  p \paren{ \yii \vert \mu_0 , \hat{\Theta}} p \paren{ \mu_0 \vert \hat{\Theta} }    \\ 
& \propto \left\{ \displaystyle \prod_{i = 1}^{M} p \paren{ \yi \vert \mu_0 , \hat{\theta}_i, \hat{\sigma}_i^2} \right\} p \paren{ \mu_0 \vert \hat{\theta}_0 } \\ 
& \propto \left\{ \displaystyle \prod_{i = 1}^{M} \mathcal{N} \paren{\yi; \mu_0, \Psiihat)} \right\}\mathcal{N} \paren{ \mu_0; m_0, \Kthetahat }.  
\end{align*}
The term $\mathcal{L}_1 = - (1/2) \log p ( \mu_0 \vert \yii, \hat{\Theta})$ may then be written as
\begin{align*}
\mathcal{L}_1 
&= - \dfrac{1}{2} \log p ( \mu_0 \vert \yii, \hat{\Theta}) \\
&= \sumi \paren{ y_i - \mu_0 }^{\intercal} \Psiihat^{-1} \paren{ y_i - \mu_0 } + \paren{ \mu_0 - m_0 }^{\intercal} \Kthetahat^{-1} \paren{ \mu_0 - m_0 } + C_1 \\
&= \sumi \mu_0^{\intercal} \Psiihat^{-1} \mu_0 - 2 \mu_0^{\intercal} \Psiihat^{-1} \yi  + \mu_0^{\intercal} \Kthetahat^{-1} \mu_0 - 2 \mu_0^{\intercal} \Kthetahat^{-1} m_0 + C_2 \\
&= \mu_0^{\intercal} \paren{ \Kthetahat^{-1} + \sumi \Psiihat^{-1} } \mu_0 - 2 \mu_0^{\intercal} \paren{ \Kthetahat^{-1} m_0 + \sumi \Psiihat^{-1} \yi} + C_2,
\end{align*}
where the constant terms are gathered into $C_1, C_2 \in \mathbb{R}$.
Identifying terms in the quadratic form with the Gaussian likelihood, we get the desired result.
\qed
\end{proof}

The maximisation step depends on the assumptions on the generative model, resulting in two versions for the EM algorithm (the E step is common to both, the branching point is here).

\paragraph{M step: different hyper-parameters}
~\par

Assuming each individual has its own set of hyper-parameters $\{ \theta_i, \sigma_i^2 \}$, the M step is given by the following procedure.
\begin{proposition}
\label{prop:M_step_diff}
Assume $p(\mu_0 \vert \yii) = \mathcal{N} \paren{\mu_0(\Ut); \mhat(\Ut), \Khat^{\Ut} }$ computed in a previous E step.
For a set of hyper-parameters $\Theta = \{ \theta_0, \thetaii, \sigmaii \}$, optimal values are given by
\begin{align*}
\hat{\Theta}
&= \argmax_{\Theta} \mathbb{E}_{\mu_0 \vert \yii} \croch{ p(\yii, \mu_0(\Ut) \vert \Theta) },
\end{align*}
\noindent inducing $M +1$ independent maximisation problems: 
\begin{align*}
\hat{\theta}_0  &= \argmax\limits_{\theta_0} \ \mathcal{L}^{\Ut} \paren{\mhat(\Ut); m_0(\Ut), \mathbf{K}_{\theta_0}^{\Ut} } , \\
( \hat{\theta}_i, \hat{\sigma}_i^2 ) &= \argmax\limits_{\theta_i, \sigma_i^2}  \ \mathcal{L}^{\ti} ( \yi; \mhat(\Ut), \Psii^{\ti} ), \ \forall i,
\end{align*}
\noindent where
\begin{equation*}
\mathcal{L}^{\Ut} \paren{ \mathbf{x}; \mathbf{m}, \mathbf{S} } = \log \mathcal{N} \paren{\mathbf{x}; \mathbf{m}, \mathbf{S} } - \dfrac{1}{2} \operatorname{Tr} \paren{ \Khat^{\Ut}  {\mathbf{S}}^{-1}}.
\end{equation*}
\end{proposition}
\begin{proof}
One simply has to distribute the conditional expectation in order to get the right likelihood to maximise, and then notice that the function can be written as a sum of $M+1$ independent (with respect to the hyper-parameters) terms. 
Moreover, by rearranging, one can observe that each independent term is the sum of a Gaussian likelihood and a correction trace term. See \Cref{proof:Proof_M_step} for details.
\qed
\end{proof}

\paragraph{M step: common hyper-parameters}
~\par

Alternatively, assuming all individuals share the same set of hyper-parameters $\{ \theta, \sigma^2 \}$, the M step is given by the following procedure.
\begin{proposition}
\label{prop:M_step_common}
Assume $p(\mu_0 \vert \yii) = \mathcal{N} \paren{\mu_0(\Ut); \mhat(\Ut), \Khat^{\Ut} }$ computed in a previous E step.
For a set of hyper-parameters  $\Theta = \{ \theta_0, \theta, \sigma^2 \}$, optimal values are given by
\begin{equation*}
\hat{\Theta} = \argmax_{\Theta} \mathbb{E}_{\mu_0 \vert \yii} \croch{ p(\yii, \mu_0(\Ut) \vert \Theta) }, 
\end{equation*}
\noindent inducing two independent maximisation problems: 
\begin{align*}
\hat{\theta}_0  &= \argmax\limits_{\theta_0} \ \mathcal{L}^{\Ut} \paren{\mhat(\Ut); m_0(\Ut), \mathbf{K}_{\theta_0}^{\Ut} } , \\
( \hat{\theta}, \hat{\sigma}^2 ) &= \argmax\limits_{\theta, \sigma^2}  \ \mathcal{L}_M ( \theta, \sigma^2 ),
\end{align*}
\noindent where 		
\begin{equation*}
\mathcal{L}_M ( \theta, \sigma^2 ) = \sumi \mathcal{L}^{\ti} ( \yi; \mhat(\Ut), \boldsymbol{\Psi}_{\theta, \sigma^2}^{\ti} ).
\end{equation*}
\end{proposition}
\begin{proof}
We use the same strategy as for \Cref{prop:M_step_diff}, see \Cref{proof:Proof_M_step} for details. \qed
\end{proof}

In both cases, explicit gradients associated with the likelihoods to maximise are available, facilitating the optimisation with gradient-based methods.

\subsection{Initialisation}
\label{sec:initialisation}

To implement the EM algorithm described above, several constants must be (appropriately) initialised:

\begin{itemize}
\item $m_0(\cdot)$, the mean parameter from the hyper-prior distribution of the process $\mu_0(\cdot)$.
A somewhat classical choice in GP is to set its value to a constant function, typically $0$ in the absence of external knowledge.
Notice that, in our multi-task framework, the influence of $m_0(\cdot)$ in hyper-posterior computation decreases as $M$ grows anyway (see  \Cref{prop:E_step}).
\item Initial values for kernel parameters $\theta_0$ and $\thetaii$. 
Those strongly depend on the chosen kernel and its properties.
We advise initiating $\theta_0$ and $\thetaii$ with close values, as a too large difference might induce nearly singular covariance matrices and result in numerical instability (typical in GPs applications).
In such pathological regime, the influence of a specific individual tends to overtake others in the calculus of $\mu_0$'s hyper-posterior distribution.
\item Initial values for the variance of the error terms $\sigmaii$. 
This choice mostly depends on the context and properties of the dataset. 
We suggest avoiding initial values with more than an order of magnitude different from the variability of data. 
In particular, a too high value might result in a model mostly capturing noise.
\end{itemize}

As a final  note, let us stress  that the EM algorithm  depends on the
initialisation and is only
guaranteed   to   converge   to   local  maxima   of   the   likelihood
function \citep{McLachlanEMAlgorithmExtensions2007}. Several strategies have been considered in the literature to
tackle this issue such as simulated annealing \citep{UedaDeterministicannealingEM1998} or repeated short runs \citep{BiernackiChoosingstartingvalues2003}. In this work, we chose the latter option.

\subsection{Pseudocode}
\label{sec:pseudo_code}

We wrap up this section with the pseudocode of the EM component of our complete algorithm, which we call \algo (standing for Multi tAsk Gaussian processes with common MeAn). The corresponding code is available at \url{https://github.com/ArthurLeroy/MAGMA}.

\begin{algorithm}
\caption{\algo: EM component}
\label{alg:algo_EM}
\begin{algorithmic}
\STATE Initialise $m_0$ and $\Theta = \acc{ \theta_0 , \thetaii , \sigmaii}$.
\WHILE{not converged} 
\STATE E step: Compute the hyper-posterior distribution
\STATE \hspace{1.1cm} $p(\mu_0 \vert \yii, \hat{\Theta}) = \mathcal{N}(\mhat, \Khat).$ 
\newline

\STATE M step: Estimate hyper-parameter by maximising
\STATE \hspace{1.1cm} $\hat{\Theta} = \argmax\limits_{\Theta} \mathbb{E}_{\mu_0 \vert \yii} \croch{ p(\yii , \mu_0 \vert \Theta) } .$
\ENDWHILE
\RETURN $\hat{\Theta}$, $\mhat$, $\Khat$.
\end{algorithmic}
\end{algorithm}

\subsection{Discussion of EM algorithms and alternatives}
\label{sec:related_work}

Let us stress that even though we focus on prediction purpose in this paper, the output of the EM algorithm already provides results on related FDA problems.
The generative model in \cite{YangSmoothingMeanCovariance2016} describes a Bayesian framework that resembles ours to smooth multiple curves simultaneously. 
However, modelling variance structure with an Inverse-Wishart process forces the use of an MCMC algorithm for inference or the introduction of a more tractable approximation in \cite{YangEfficientBayesianhierarchical2017a}.
One can think of the learning through \algo and applying a single task GP regression on each individual as an \textit{empirical Bayes} counterpart to their approach.
Meanwhile, $\mu_0$'s hyper-posterior distribution also provides the probabilistic estimation of a mean curve from a set of functional data.
The closest method to our approach can be found in \cite{ShiGaussianProcessFunctional2007} and the following book \cite{ShiGaussianProcessRegression2011}.
The authors also work in the context of a multi-task GP model, and one can retrieve the idea of defining a mean function $\mu_0$ to overcome the weaknesses of classic GPs in making predictions far from observed data.
However, since their model uses B-splines to estimate this mean function, the method only works if all individuals share the same grid of observations, and does not account for uncertainty over $\mu_0$.

\section{Prediction}
\label{sec:prediction}

Once the hyper-parameters of the model have been learned, we can focus on our main goal: prediction for new individuals at unobserved timestamps.
Since $\hat{\Theta}$ is known and for the sake of concision, we omit conditioning on $\hat{\Theta}$ in the sequel.
Note there are two cases for prediction \citep[referred to as \emph{Type I} and \emph{Type II} in][Section 3.2.1]{ShiGaussianProcessFunction2014a}, depending on whether we observe some data or not for any new individual we wish to predict on.
We denote by the index $*$ a new individual for whom we want to make a prediction, say at timestamps $\tpred$.
If there are no available data for this individual, we have no $*$-specific information, and the prediction is merely given by $p(\mu_0(\tpred) \vert \yii)$.
This quantity may be considered as the 'generic' (or \emph{Type II}) prediction according to the trained model, and only informs us through the mean process.
Computing $p(\mu_0(\tpred) \vert \yii)$ is also one of the steps leading to the prediction for a partially observed new individual (\emph{Type I}).
The latter being the most compelling case, we consider \emph{Type II} prediction as a particular case of the full \emph{Type I} procedure, described below.
\newline

If we observe $\acc{ \tst, \yst (\tst) }$ for the new individual, the multi-task GP prediction is obtained in our model by computing the posterior distribution $p(\yst (\tpred) \vert \yst(\tst), \yii )$. 
Note that the conditioning is taken over $\yst(\tst)$, as for any GP regression, but also on $\yii$, which is specific to our multi-task setting.
The procedure for computing this distribution requires to successively complete the following steps:
\begin{enumerate}
	\item choose a grid of prediction $\tpred$ and define the pooled vector of timestamps
	$\tpst$,
	\item compute the hyper-posterior distribution of $\mu_0$ at $\tpst$: $p(\mu_0(\tpst) \vert \yii )$,
	\item compute the multi-task prior distribution $p(\yst (\tpst) \vert \yii )$,
	\item compute hyper-parameters $\theta_*$ associated with the new individual (optional),
	\item compute the multi-task posterior distribution: $p(\yst (\tpred) \vert \yst(\tst), \yii )$.
\end{enumerate}

\subsection{Posterior inference on the mean process}
\label{sec:Posterior_mu0}

As mentioned above, we observed a new individual at timestamps $\tst$. 
The GP regression consists in arbitrarily choosing a vector $\tpred$ of timestamps for which we aim at making predictions.
Then, we define new notation for the pooled vector of timestamps $\tpst = 
\begin{bmatrix}
\tpred \\
\tst
\end{bmatrix}$, which will serve as a working grid to define the prior and posterior distributions involved in the prediction process.
One can note that, although not mandatory in theory, it is often a good idea to include the observed timestamps of training individuals, $\Ut$, within $\tpst$ since they match locations that contain information for the mean process to 'help' the prediction. 
In particular, if $\tpst = \Ut$, the computation of $\mu_0$'s hyper-posterior distribution is not necessary since $p(\mu_0(\Ut) \vert \yii)$ has previously been obtained from the EM algorithm. 
However, in general, it is necessary to compute the hyper-posterior $p(\mu_0(\tpst) \vert \yii)$ at the new timestamps.
The idea remains similar to the E step aforementioned, and we obtain the following result.
\begin{proposition}
\label{prop:post_mu}
\noindent Let $\tpst$ be a vector of timestamps of size $\tilde{N}$.
The hyper-posterior distribution of $\mu_0$ remains Gaussian: 
\begin{equation*}
p\paren{\mu_0(\tpst) \vert \yii} = \mathcal{N} \paren{\mu_0(\tpst); \mhat(\tpst), \Khat_{*}^{p}},
\end{equation*}
with: 
\begin{itemize}
	\item $\Khat_{*}^{p} = \paren{\tilde{\mathbf{K}}^{-1} + \sum\limits_{i = 1}^{M}{\tilde{\boldsymbol{\Psi}}_i}^{-1}}^{-1}$,
	\item $\mhat(\tpst) = \Khat_{*}^{p} \paren{ \tilde{\mathbf{K}}^{-1} m_0\paren{\tpst} + \sum\limits_{i = 1}^{M}{\tilde{\boldsymbol{\Psi}}_i^{-1} \tilde{\mathbf{y}}_i } }$,
\end{itemize}
where we used the shortening notation: 
\begin{itemize}
	\item $\tilde{\mathbf{K}} = k_{\hat{\theta}_0} \paren{\tpst, \tpst}$ ($\tilde{N}\times\tilde{N}$ matrix),
	\item $\tilde{\mathbf{y}}_i = \paren{\mathds{1}_{ [t \in \ti ]} \times y_i(t)}_{t \in \tpst}$ ($\tilde{N}$-size vector),
	\item $\tilde{\boldsymbol{\Psi}}_i = \croch{ \mathds{1}_{ [t, t' \in \ti]} \times \psiihat \paren{t, t'} }_{t, t' \in \tpst}$ ($\tilde{N}\times\tilde{N}$ matrix).
\end{itemize}
\end{proposition}
\begin{proof}
The sketch of the proof is similar to \Cref{prop:E_step} in the E step. The only technicality consists in dealing carefully with the dimensions of vectors and matrices involved, and whenever relevant, to define augmented versions of $\yi$ and $\Psiihat$ with 0 elements at unobserved timestamps' position for the $i$-th individual.
Note that if we pick a vector $\tpst$ including only some of the timestamps from $\ti$, information coming from $y_i$ at the remaining timestamps is ignored. We defer details to \Cref{proof:E_step}. \qed
\end{proof}

\subsection{Computing the multi-task prior distribution}
\label{sec:prior_new_indiv}

According to our generative model, given the mean process, any new individual $*$ is modelled as:
\begin{equation*}
y_*(\cdot) \vert \mu_0(\cdot) \sim \mathcal{GP} \paren{ \mu_0(\cdot), \boldsymbol{\Psi}_{\theta_*, \sigma_*^2}( \cdot, \cdot) }. 
\end{equation*}

Therefore, for any finite-dimensional vector of timestamps, and in particular for $\tpst$, $p(y_*(\tpst) \vert \mu_0(\tpst))$ is a multivariate Gaussian. Moreover, from this distribution and $\mu_0$'s hyper-posterior, we can figure out the multi-task prior distribution over $y_*(\tpst)$, defined as below.
\begin{proposition}
\label{prop:integrate_mu}
For any set of timestamps $\tpst$, the multi-task prior distribution of $y_*$ is given by 
\begin{equation}
\label{eq:prior}
p(y_*(\tpst) \vert \yii) = \mathcal{N} \paren{ y_*(\tpst); \mhat(\tpst), \Khat_*^p + \boldsymbol{\Psi}_{\theta_*, \sigma_*^2}^{\tpst}}.
\end{equation}
\end{proposition}
\begin{proof}
To compute this prior, we need to integrate out the mean process $\mu_0$ in $p(y_* \vert \mu_0, \yii)$, whereas the multi-task aspect remains through the conditioning over $\yii$.
We omit the writing of timestamps, by using the simplified notation $\mu_0$ and $y_*$ instead of $\mu_0(\tpst)$ and $y_*(\tpst)$, respectively. 
We first use the assumption that $\{ y_i \vert \mu_0 \}_{i \in \{ 1,\dots,M \}} \indep y_* \vert \mu_0$, \emph{i.e.}, the individuals are independent conditionally to $\mu_0$. Then, one can notice that the two distributions involved within the integral are Gaussian, which leads to the explicit Gaussian target distribution after integration. 
\begin{align*} p(y_* \vert \yii)
&= \int\limits p \paren{y_*, \mu_0 \vert \yii} \dif \mu_0 \\
&= \int\limits p \paren{y_*\vert \mu_0, \yii) p(\mu_0 \vert \yii} \dif \mu_0 \\
&= \int\limits \underbrace{p \paren{y_*\vert \mu_0)}}_{\mathcal{N} \paren{y_*; \mu_0, \boldsymbol{\Psi}_{\theta_*, \sigma_*^2}^{\tpst}} } \underbrace{p(\mu_0 \vert \yii)}_{\mathcal{N} \paren{\mu_0; \mhat, \Khat_*^p} } \dif \mu_0.
\end{align*}

This convolution of two Gaussians remains Gaussian \citep[][Chapter 2.3.3]{BishopPatternrecognitionmachine2006}.
The mean parameter is then given by
\begin{align*}
\mathbb{E}_{y_*\vert \yii}\croch{y_*} 
&= \int y_* \ p \paren{ y_* \vert \yii } \dif y_* \\
&= \int y_* \int p \paren{y_*\vert \mu_0} p(\mu_0 \vert \yii) \dif \mu_0 \dif y_* \\ 
&= \int \paren{ \int y_* p \paren{y_*\vert \mu_0} \dif y_*} p(\mu_0 \vert \yii) \dif \mu_0  \\ 
&= \int \mathbb{E}_{y_* \vert \mu_0} \croch{ y_* } p(\mu_0 \vert \yii) \dif \mu_0  \\ 
&= \mathbb{E}_{\mu_0 \vert \yii} \croch{ \mathbb{E}_{y_* \vert \mu_0} \croch{ y_* } } \\
&= \mathbb{E}_{\mu_0 \vert \yii} \croch{ \mu_0 } \\
&= \mhat.
\end{align*}

Following the same idea, the second-order moment is given by
\begin{align*}
\mathbb{E}_{y_*\vert \yii}\croch{y_*^2} 
&= \mathbb{E}_{\mu_0 \vert \yii} \croch{ \mathbb{E}_{y_* \vert \mu_0} \croch{ y_*^2 } } \\
&= \mathbb{E}_{\mu_0 \vert \yii} \croch{ \mathbb{V}_{y_* \vert \mu_0} \croch{y_*} + \mathbb{E}_{y_* \vert \mu_0} \croch{ y_* }^2 } \\
&= \boldsymbol{\Psi}_{\theta_*, \sigma_*^2} + \mathbb{E}_{\mu_0 \vert \yii} \croch{ \mu_0^2 } \\
&= \boldsymbol{\Psi}_{\theta_*, \sigma_*^2} + \mathbb{V}_{\mu_0 \vert \yii} \croch{\mu_0} + \mathbb{E}_{\mu_0 \vert \yii} \croch{ \mu_0 }^2 \\
&= \boldsymbol{\Psi}_{\theta_*, \sigma_*^2} + \Khat + \mhat^2,
\end{align*}

\noindent hence 
\begin{align*}
\mathbb{V}_{y_*\vert \yii}\croch{y_*} 
&= \mathbb{E}_{y_*\vert \yii}\croch{y_*^2}  - \mathbb{E}_{y_*\vert \yii}\croch{y_*}^2 \\
&= \boldsymbol{\Psi}_{\theta_*, \sigma_*^2} + \Khat + \mhat^2 - \mhat^2 \\
&= \boldsymbol{\Psi}_{\theta_*, \sigma_*^2} + \Khat.
\end{align*}
\qed
\end{proof}

Note that the process $y_*(\cdot) \vert \yii$ is not strictly a GP, although its finite-dimensional evaluation \eqref{eq:prior} remains Gaussian.
The covariance structure cannot be expressed as a kernel that could be directly evaluated at any timestamps: the process is known as a \emph{degenerated GP}.
In practice however, this does not bear much consequence as any arbitrary vector of timestamps $\tau$ can be chosen at first, and computing hyper-posterior $p(\mu_0(\tau) \vert \yii)$ still yields to the Gaussian distribution $p(y_*(\tau) \vert \yii)$ as above. 
For the sake of simplicity, we now rename the covariance matrix of the multi-task prior distribution:
\begin{equation*}
\Khat_*^p + \boldsymbol{\Psi}_{\theta_*, \sigma_*^2}^{\tpst} = \boldsymbol{\Gamma}_*^p = 				   
\begin{pmatrix}
	\boldsymbol{\Gamma}_{pp} & \boldsymbol{\Gamma}_{p*} \\
	\boldsymbol{\Gamma}_{*p} & \boldsymbol{\Gamma}_{**}
\end{pmatrix},
\end{equation*}
\noindent where the indices in the blocks of the matrix correspond to the associated timestamps $\tpred$ and $\tst$.

\subsection{Learning the new hyper-parameters}
\label{sec:Learning_new_hp}

When we collect data points for a new individual, as in the single-task GPs setting, we would need to learn the hyper-parameters of its covariance kernel before making predictions. 
A salient fact in our multi-task approach is that we consider this step being part of the prediction process, for two main reasons.
First, the model is already trained for individuals $i = 1,\dots, M$, and this training is independent of the future individual $*$ or the choice of prediction timestamps. 
Since learning these new hyper-parameters requires knowledge of $\mu(\tpst)$ and thus of the prediction timestamps, we cannot compute them beforehand.
Second, learning these hyper-parameters with the \emph{empirical Bayes} approach only requires maximisation of a Gaussian likelihood which is negligible in computing time compared to the previous EM algorithm. 
As for single-task GP, we have the following estimates for hyper-parameters:
\begin{align*}
\hat{\Theta}_* 
&= \argmax_{\Theta_*} p(y_*(\tst) \vert \yii, \Theta_*) \\
&= \argmax_{\Theta_*} \mathcal{N}\paren{ y_*(\tst); \mhat(\tst), \boldsymbol{\Gamma}_{**}^{\Theta_*} }.
\end{align*}

Note that this step is optional depending on the modelling assumption: in the common hyper-parameters model (i.e. $(\theta, \sigma^2) = (\theta_i, \sigma_i^2), \forall i \in \I$), any new individual will also share the same hyper-parameters and we already have $\hat{\Theta}_* = (\hat{\theta}_*, \hat{\sigma}_*^2) = (\hat{\theta}, \hat{\sigma}^2)$ from the EM algorithm.

\subsection{Prediction}
\label{sec:GP_pred}

We can rewrite the multi-task prior distribution, by separating observed and prediction timestamps, as: 
\begin{align*}
p( y_*(\tpst) \vert \yii) 
&= p( y_*(\tpred), y_*(\tst) \vert \yii) \\
&= \mathcal{N} \paren{y_*(\tpst); \mhat(\tpst), \boldsymbol{\Gamma}_*^p } \\
&=  \mathcal{N} \paren{
\begin{bmatrix}
y_*(\tpred) \\
y_*(\tst)
\end{bmatrix};
\begin{bmatrix}
\mhat(\tpred) \\
\mhat(\tst)
\end{bmatrix}, 
\begin{pmatrix}
\boldsymbol{\Gamma}_{pp} & \boldsymbol{\Gamma}_{p*} \\
\boldsymbol{\Gamma}_{*p} & \boldsymbol{\Gamma}_{**}
\end{pmatrix} }.
\end{align*}

\noindent As usual, the conditional distribution remains Gaussian, and the multi-task posterior distribution is given by: 
\begin{equation*}
p( y_*(\tpred) \vert y_*(\tst) , \yii) = \mathcal{N} \paren{ y_*(\tpred); \hat{\mu}_0^p, \hat{\boldsymbol{\Gamma}}^p },  
\end{equation*}

\noindent where: 

\begin{itemize}
\item $\hat{\mu}_0^p = \mhat(\tpred) + \boldsymbol{\Gamma}_{p*} \boldsymbol{\Gamma}_{**}^{-1} \paren{ y_*(\tst) - \mhat(\tst)},$
\item $\hat{\boldsymbol{\Gamma}}^p = \boldsymbol{\Gamma}_{pp} - \boldsymbol{\Gamma}_{p*} \boldsymbol{\Gamma}_{**}^{-1} \boldsymbol{\Gamma}_{*p}.$
\end{itemize}

Although this predictive distribution presents a formulation nicely analogous to standard GPs, let us emphasise on the terms $\mhat(\tpst)$ and $\boldsymbol{\Gamma}_*^p$, which embed crucial information from training individuals for the mean prediction to be more relevant even in far from the observed points $y_*(\tst)$.

\section{Complexity analysis for training and prediction}
\label{sec:complexity}

Computational complexity is of paramount importance in GPs as it quickly scales with large datasets.
The classical cost to train a GP is $\mathcal{O}(N^3)$, and $\mathcal{O}(N^2)$ for prediction \citep{RasmussenGaussianprocessesmachine2006a} where $N$ is the number of data points (although there exist various sparse approximations, see \Cref{sec:conclusion} for references).
Moreover, multi-task GP models lying on LMC approaches typically present a complexity of $\mathcal{O}(M^3 N^3)$ in training, which can be diminished when using sparse approximations \citep{AlvarezComputationallyEfficientConvolved2011a}. 
As detailed below, our model reaches a reduction to $\mathcal{O}((M + 1) N^3)$ for the training complexity in a similar context (common grid of timestamps for all individuals), without using any sparse approximation. 
\newline

More specifically, since \algo uses information from $M$ individuals, each of them providing $N_i$ observations, these quantities determine the overall complexity of the algorithm. 
If we recall that $N$ is the number of distinct timestamps (\emph{i.e.} $N \leq \sumi N_i$), the training complexity is $\mathcal{O} \paren{M \times N_i^3 + N^3 }$ (\emph{i.e.} the complexity of each EM iteration). 
As usual with GPs, the cubic costs come from the inversion of the corresponding matrices, and here, the constant is proportional to the number of iterations of the EM algorithm. 
The dominating term in this expression depends on the values of $M$, relatively to $N$.
For a large number of individuals with many common timestamps ($MN_i \gtrsim N$), the first term dominates.
For diverse timestamps among individuals ($MN_i \lesssim N$), the second term becomes the primary burden, as in any GP problem.
During the prediction step, the re-computation of $\mu_0$'s hyper-posterior implies the inversion of a $\tilde{N} \times \tilde{N}$ (dimension of $\tpst$) which has a $\mathcal{O}(\tilde{N}^3)$ complexity while the new hyper-parameters estimation's cost is $\mathcal{O}(N_*^3)$. 
In practice, the most computationally-expensive steps can be performed in advance to allow for quick on-the-fly prediction when collecting new data. 
If we observe the training dataset once and pre-compute the hyper-posterior of $\mu_0$ on a fine grid on which to predict later, the immediate computational cost for each new individual is identical to the one of the single-task GP regression.
\section{Experimental results}
\label{sec:exp}

We evaluate our \algo algorithm on synthetic data and two real datasets. 
The classical GP regression on single tasks separately is used as the baseline alternative for predictions.
While it is not expected to perform well on the dataset used, the comparison highlights the interest of multi-task approaches.
To our knowledge, the only alternative to \algo is the GPFDA algorithm from \cite{ShiGaussianProcessFunctional2007,ShiGaussianProcessRegression2011}, described in \Cref{sec:related_work}, and the associated R package \emph{GPFDA}, which is applied during the experiments. 
Throughout the section, the standard \emph{Exponentiated Quadratic} kernel (see \Cref{eq:kernel}) is used both for simulating the data and for modelling the covariance structures in the three algorithms. 
Hence, each kernel is associated with $\theta = \{ v, \ell \}, \ v, \ell \in \mathbb{R}^{+}$, a set of variance and length-scale hyper-parameters, respectively. 
Each simulated dataset has been drawn from the sampling scheme below:
\begin{enumerate}
	\item Draw a random working grid $\Ut \subset \croch{0,10}$ of $N = 200$ timestamps, and a number $M$ of individuals.
	\item Define a prior mean function : $m_0(t) = at + b, \ \forall t \in \Ut$, where $a \in \croch{-2, 2}$ and $b \in \croch{0, 10}$ are drawn uniformly.
	\item Draw hyper-parameters uniformly for $\mu_0$'s kernel : $\theta_0 = \{ v_0, \ell_0 \}$, where $v_0 \in \croch{1, \exp(5) }$ and $\ell_0 \in \croch{1, \exp(2)}$.
	\item Draw $\mu_0 (\Ut) \sim \mathcal{N} \paren{m_0(\Ut), \mathbf{K}_{\theta_0}^{\Ut}}$.
	\item $\forall i \in \I$, draw $v_i \in \croch{1, \exp(5) }$, $\ell_i \in \croch{1, \exp(2)}$, and $\sigma_i^2 \in \croch{0, 1}$ uniformly.
	\item $\forall i \in \I$, draw a subset $\ti \subset \Ut$ of $N_i = 30$ timestamps uniformly, and draw $\yi \sim \mathcal{N} \paren{\mu_0(\ti), \Psii^{\ti}}$.
\end{enumerate}

This procedure provides a synthetic dataset $\acc{\ti, \yi}_i$, and its associated mean process $\mu_0(\Ut)$. 
Those quantities are used to train the model, make predictions with each algorithm, and then compute errors in $\mu_0$ estimation and forecasts.
We recall that the \algo algorithm enables two different settings depending on the model's assumption over hyper-parameters (HP), and we refer to them as \emph{Common HP} and \emph{Different HP} in the following. 
In order to test these two contexts, differentiated datasets have been generated, by drawing \emph{Common HP data} or \emph{Different HP data} for each individual at step 5.
We previously presented the idea of the model used in GPFDA, and, although the algorithm has many features (in particular about the type and number of input variables), it is not yet usable when timestamps are different among individuals. 
Therefore, two frameworks are considered, \emph{Common grid} and \emph{Uncommon grid}, to take this specification into account.
Thus, the comparison between the different methods can only be performed on data generated under the settings \emph{Common HP} and \emph{Common grid}, and the effect of those different settings on \algo is analysed separately. 
Moreover, the initialisation for the prior mean function, $m_0(\cdot)$, is set to be constant, equal to $0$ for each algorithm. 
Except in some experiments, where the influence of the number of individuals is analysed, the generic value is $M = 20$.
In the case of prediction on unobserved timestamps for a new individual, the first $20$ data points are used as observations, and the remaining $10$ are taken as test values.
Optimisation of the hyper-parameters is performed by likelihood maximisation, using the L-BFGS-B algorithm \citep{NocedalUpdatingquasiNewtonmatrices1980, MoralesRemarkalgorithmLBFGSB2011} in all methods. 
The convergence criterion for all algorithms is reached if the difference of log-likelihood between two iterations is lower than $10^{-2}$.
In general, the EM algorithm in \algo converges in a few iterations, typically fewer than 5 with the \emph{Common HP} setting, and rarely more than 15 even with the \emph{Different HP} setting.

\subsection{Illustration on a simple example}
\label{sec:simu_illustration}

\begin{figure*}
    \begin{center}
       \includegraphics[width=\textwidth]{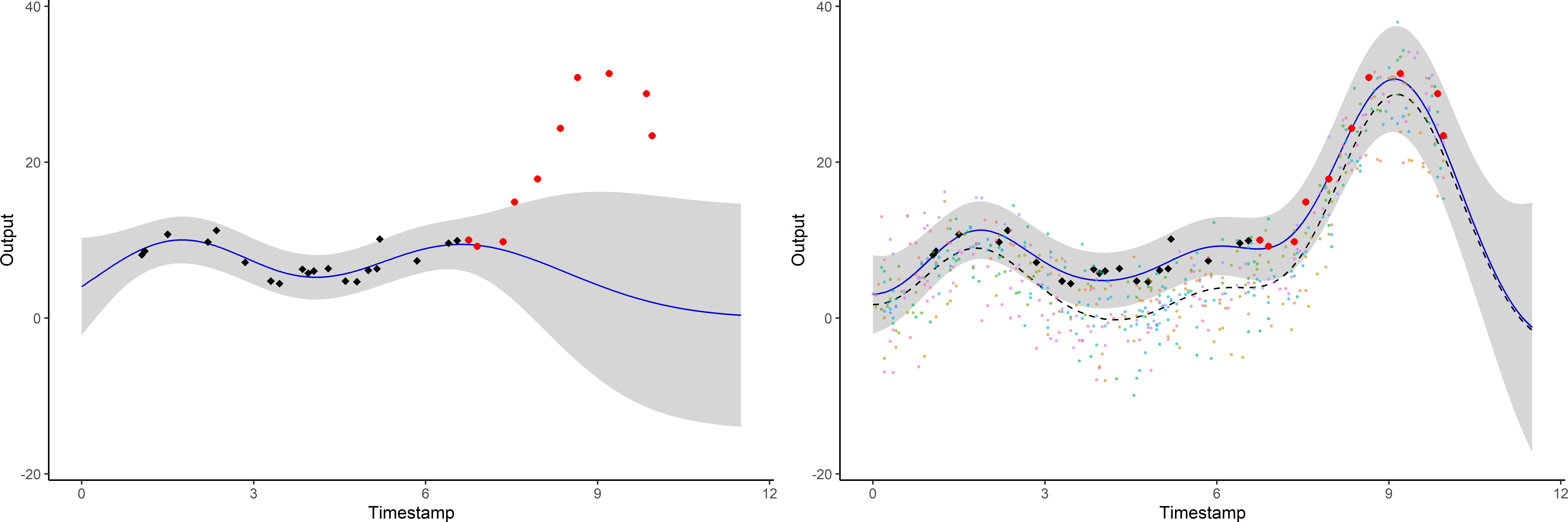}
       \caption{Prediction curves (blue) of a new individual with associated 95\% credible intervals (grey) for GP regression (left) and \algo (right). The dashed line represents the mean function $\hat{m}_0$, from the hyper-posterior $p(\mu_0 \vert \yii)$. Observed data points are in black, and testing data points are in red. The colourful backward points are the observations from the training dataset, each colour corresponding to a different individual.} 
		\label{fig:example}  
    \end{center}
\end{figure*}

To illustrate the multi-task approach of \algo, \Cref{fig:example} displays a comparison between standard GP regression and \algo on a simple example, from a dataset simulated according to the scheme above and using the \emph{Uncommon grid}/\emph{Common HP} setting. 
Given the observed data (in black), values on a thin grid of unobserved timestamps are predicted and compared, in particular, with the true test values (in red).
As expected, the GP regression provides a good fit close to the data points and then dives rapidly to the prior 0 with increasing uncertainty. 
Conversely, although the initialisation for the prior mean is $0$ in \algo as well, the hyper-posterior distribution of $\mu_0$ (dashed line) is estimated thanks to all individuals in the training dataset. 
This process acts as an informed prior helping GP prediction for the new individual, even far from its own observations.
More precisely, 3 phases can be distinguished according to the level of information coming from the data: in the first one, close to the observed data ($t \in \croch{1,7}$), the two processes behave similarly, except for a slight increase in the variance for \algo, which is logical since the prediction also takes uncertainty over $\mu_0$ into account (see \Cref{eq:prior});
in the second one, on intervals of unobserved timestamps containing data points from the training dataset ($t \in \croch{0,1} \cup \croch{7,10}$), the prediction is guided by the information coming from other individuals through $\mu_0$.
In this context, the mean trajectory remains coherent and the uncertainty increases only slightly. 
In the third phase, where no observations are available, neither from the new individual nor from the training dataset ($t \in \croch{10,12}$), the prediction behaves as expected, with a slow drifting to the prior mean 0, with highly increasing variance. 
Overall, the multi-task framework provides reliable probabilistic predictions on a wider range of timestamps, potentially outside of the usual scope of GPs.


\subsection{Performance comparison on simulated datasets}
\label{sec:simu_comparison}

\begin{table*}
\begin{center}
\caption{Average MSE (standard deviation) and average $CIC_{95}$ (standard deviation) on 100 runs for GP, GPFDA and \algo. ($\star$ : 99.6 (2.8), the measure of incertitude from the GPFDA package is not a genuine credible interval)}
\label{tab:compare_algo}
\begin{tabular}{c|cc|cc|}
\cline{2-5}
                            & \multicolumn{2}{c|}{Prediction}             & \multicolumn{2}{c|}{Estimation $\mu_0$}                 \\
                            & MSE                  & $CIC_{95}$                   & MSE                      & $CIC_{95}$                     \\ \hline
\multicolumn{1}{|c|}{\algo}  & \textbf{18.7 (31.4)} & \textbf{93.8 (13.5)} & \textbf{1.3 (2)}         & \textbf{94.3 (11.3)}     \\
\multicolumn{1}{|c|}{GPFDA} & 31.8 (49.4)          & 90.4 (18.1)          & 2.4 (3.6)                & $\star$                     \\
\multicolumn{1}{|c|}{GP}    & 87.5 (151.9)         & 74.0 (32.7)          & \cellcolor[HTML]{9B9B9B} & \cellcolor[HTML]{9B9B9B} \\ \hline
\end{tabular}
\end{center}
\end{table*}

\begin{figure*}
    \begin{center}
       \includegraphics[width= \textwidth]{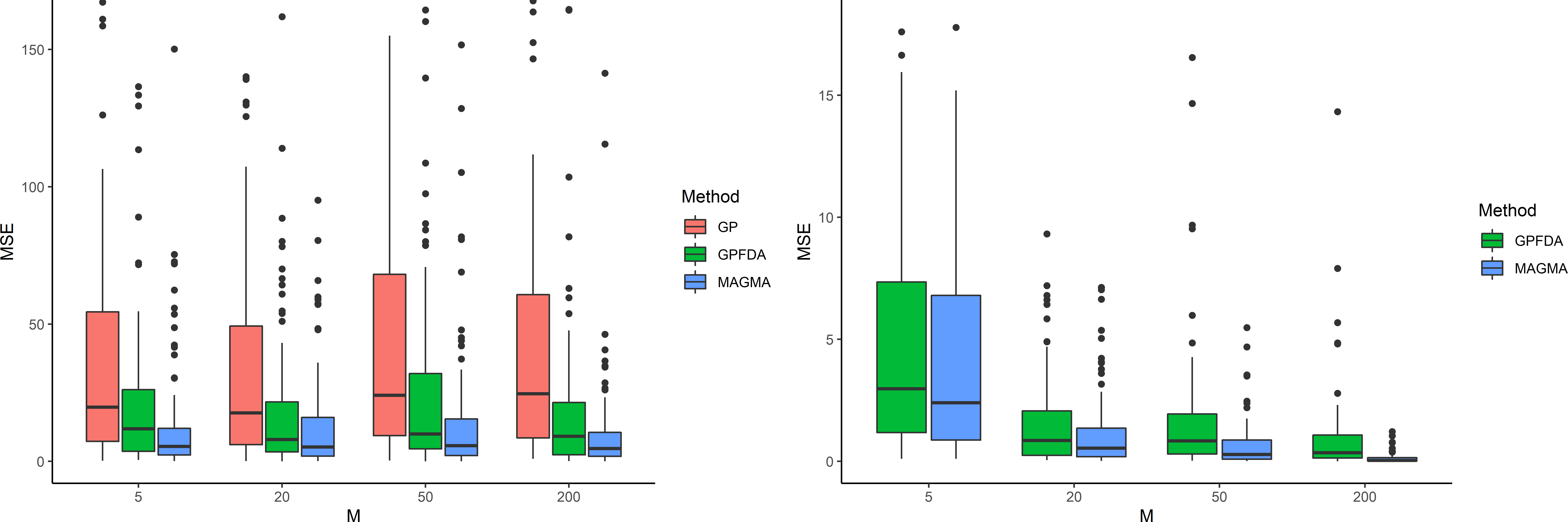}
       \caption{MSE with respect to the number $M$ of training individuals (boxplots are displayed from 100 runs in each case). \emph{Left}: prediction error on 10 testing points. \emph{Right}: estimation error of the true mean process $\mu_0$.}
		\label{fig:varying_M}  
    \end{center}
\end{figure*}

\begin{figure}
    \begin{center}
       \includegraphics[width= .5\textwidth]{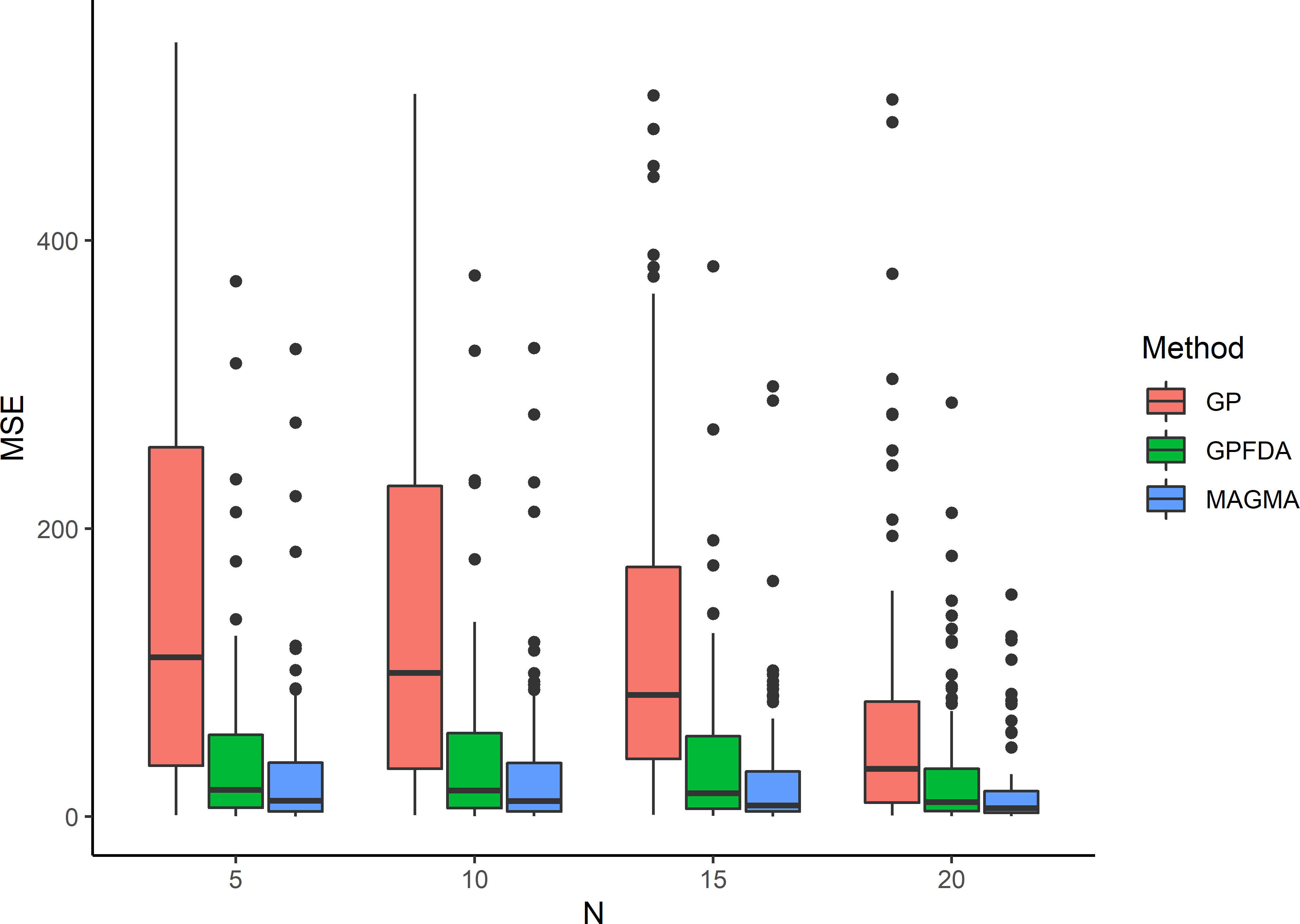}
       \caption{MSE prediction error on the 10 last testing points with respect to the increasing number N of observed timestamps, among the first 20 points (boxplots are displayed from 100 runs in each case).}
		\label{fig:varying_N}  
    \end{center}
\end{figure}

We confront the performance of \algo to alternatives in several situations and for different datasets. 
In the first place, the classical GP regression (GP), GPFDA and \algo are compared through their performance in prediction and estimation of the true mean process $\mu_0$.
In the prediction context, the performances are evaluated according to the following indicators: 
\begin{itemize}
\item the mean squared error (MSE) which compares the predicted values to the true test values of the 10 last timestamps: 
$$ \dfrac{1}{10} \sum\limits_{k = 21}^{30} \paren{ y_*^{\operatorname{pred}} (t_*^k) - y_*^{\operatorname{true}} (t_*^k) }^2 , $$
\item the $CI_{95}$ coverage ($CIC_{95}$), i.e. the percentage of unobserved data points effectively lying within the 95\% credible interval defined from the predictive posterior distribution $p(y_*(\tpred) \vert y_*(\tst), \yii)$: 
$$ 100 \times \dfrac{1}{10} \sum\limits_{k = 21}^{30} \mathds{1}_{ \{ y_*^{\operatorname{true}}(t_*^k) \in \ CI_{95} \} }.$$

\end{itemize}

The $CIC_{95}$ provides insights on the reliability of the predictive variance and should be as close to the value 95\% as possible. 
Other values would indicate a tendency to underestimate or overestimate the uncertainty. 
Let us recall that GPFDA uses B-splines to estimate the mean process and does not account for uncertainty, contrarily to a probabilistic framework as \algo.
However, a measure of uncertainty based on an empirical variance estimated from training curves is proposed \citep[see][Section 3.2.1]{ShiGaussianProcessFunction2014a}.
In practice, this measure constantly overestimates the true variance, and their 95\% empirical interval coverage is generally equal or close to 100\%.
\newline

In the estimation context, the performances are evaluated thanks to another MSE, which compares the estimations to the true values of $\mu_0$ at all timestamps: 
$$ \dfrac{1}{M} \sumi \dfrac{1}{N_i} \sum\limits_{k = 1}^{N_i} \paren{ \mu_0^{\operatorname{pred}} (t_i^k) - \mu_0^{\operatorname{true}} (t_i^k) }^2 .$$

\Cref{tab:compare_algo} presents the results obtained over 100 datasets, where the models are trained on $M = 20$ individuals, each of them observed on $N = 30$ common timestamps. 
As expected, both multi-task methods lead to better results than GP. 
However, \algo outperforms GPFDA, both in the estimation of $\mu_0$ and in predictive performance. 
In terms of error as well as in uncertainty quantification, \algo provides more accurate results, in particular with a $CI_{95}$ coverage close to the 95\% expected value. 
Each method presents a quite high standard deviation for MSE in prediction, which is due to some datasets with particularly difficult values to predict, although most of the cases lead to small errors. 
This behaviour is reasonably expected since such 10-timestamps-ahead forecasts might sometimes be tricky.
It can also be noticed on \Cref{fig:varying_M} that \algo consistently provides lower errors as well as less pathological behaviour, as it may sometimes occur with the B-splines modelling used in GPFDA.
\newline

To highlight the effect of the number of individuals $M$ on the performance, \Cref{fig:varying_M} provides the same 100 runs trial as previously, for different values of $M$. 
The boxplots exhibit, for each method, the behaviour of the prediction and estimation MSE as information is added in the training dataset.
Let us mention the absence of discernible changes as soon as $M > 200$.
As expected, we notice on the right panel that adding information from new individuals improves the estimation of $\mu_0$, leading to shallow errors for high values of $M$, in particular for \algo. 
Meanwhile, the left panel exhibits reasonably unchanged prediction performance with respect to the values of $M$, excepted for some random fluctuations.
This property is expected for GP regression since no external information is used from the training dataset in this context.
For both multi-tasks algorithms though, the estimation of $\mu_0$ improves the prediction by one order of magnitude below the typical errors, even with only a few training individuals.
Furthermore, since a new individual behaves independently through $f_*$, it is natural for a 10-points-ahead forecast to present intrinsic variations, despite an adequate estimation of the shared mean process.
\newline

To illustrate the advantage of multi-task methods, even for $M = 20$, we display on  \Cref{fig:varying_N} the evolution of MSE according to the number of timestamps $N$ that are assumed to be observed for the new individual on which we make predictions.
These predictions remain computed on the last 10 timestamps, although in this experiment, we only observe the first 5, 10, 15, or 20 timestamps, in order to change the volume of information and the distance from training observations to targets. 
We observe on \Cref{fig:varying_N} that, as expected in a GP framework, the closer observations are to targets, the better the results.
However, for multi-tasks approaches and in particular for \algo, the prediction remains consistently adequate even with few observations. 
Once more, sharing information across individuals significantly helps the prediction, even for small values of $M$ or few observed data. 

\subsection{\algo 's specific settings}
\label{sec:simu_settings}

As we previously discussed, different settings are available for \algo according to the nature of data and the model hypotheses. 
First, the \emph{Common grid} setting corresponds to cases where all individuals share the same timestamps, whereas \emph{Uncommon grid} is used otherwise.
Moreover, \algo enables to consider identical hyper-parameters for all individuals or specific ones, as previously discussed in \Cref{sec:model_hypo}.
To evaluate the effect of the different settings, performances in prediction and $\mu_0$'s estimation are evaluated in the following cases in \Cref{tab:settings_mtgp}: 
\begin{itemize}
	\item \emph{Common HP}, when data are simulated with a common set of hyper-parameters for all individuals, and \Cref{prop:M_step_common} is used for inference in \algo,
	\item \emph{Different HP}, when data are simulated with its own set of hyper-parameters for each individual, and \Cref{prop:M_step_diff} is used for inference in \algo,
	\item \emph{Common HP on different HP data}, when data are simulated with its own set of hyper-parameters for each individual, and \Cref{prop:M_step_common} is used for inference in \algo.
\end{itemize}

Note that the first line of the table (\emph{Common grid / Common HP}) of \Cref{tab:settings_mtgp} is identical to the corresponding results in \Cref{tab:compare_algo}, providing reference values, significantly better than for other methods.
The results obtained in \Cref{tab:settings_mtgp} indicate that the \algo performance is not significantly altered by the settings used or the nature of the simulated data.
To confirm the robustness of the method, the setting \emph{Common HP} was applied to data generated by drawing different values of hyper-parameters for each individual (\emph{Different HP data}). 
In this case, performances in prediction and estimation of $\mu_0$ are slightly deteriorated, although \algo still provides quite reliable forecasts.
This experience also highlights a particularity of the \emph{Different HP} setting: looking at the estimation of $\mu_0$ performance, we observe a significant decrease in the $CI_{95}$ coverage, due to numerical instability in some pathological cases. 
Numerical issues, in particular during matrix inversions, are classical problems in the GP literature and, because of the potentially large number of different hyper-parameters to train, the probability for at least one of them to lead to a nearly singular matrix increases.
In this case, one individual might overwhelm others in the calculus of $\mu_0$'s hyper-posterior (see \Cref{prop:post_mu}), and thus lead to an underestimated posterior variance. 
This problem does not occur in the \emph{Common HP} settings, since sharing the same hyper-parameters prevents the associated covariance matrices from running over each other.
Thus, except if one specifically wants to smooth multiple curves presenting really different behaviours,  keeping \emph{Common HP} as a default setting appears as a reasonable choice.
Let us notice that the estimation of $\mu_0$ is slightly better for common than for uncommon grid since the estimation problem on the union of different timestamps is generally more difficult. 
However, this feature only depends on the nature of data.

\begin{table*}
\begin{center}
\caption{Average MSE (standard deviation) and average $CIC_{95}$ (standard deviation) on 100 runs for the different settings of \algo.}
\label{tab:settings_mtgp}
\begin{tabular}{cl|ll|ll|}
\cline{3-6}
\multicolumn{1}{l}{}                                                                                             &               & \multicolumn{2}{c|}{Prediction}                     & \multicolumn{2}{c|}{Estimation of $\mu_0$}                 \\
\multicolumn{1}{l}{}                                                                                             &               & \multicolumn{1}{c}{MSE} & \multicolumn{1}{c|}{$CIC_{95}$} & \multicolumn{1}{c}{MSE} & \multicolumn{1}{c|}{$CIC_{95}$} \\ \hline
\multicolumn{1}{|c|}{\multirow{2}{*}{Common HP}}                                                                 & Common grid   & 18.7 (31.4)             & 93.8 (13.5)               & 1.3 (2)                 & 94.3 (11.3)               \\
\multicolumn{1}{|c|}{}                                                                                           & Uncommon grid & 19.2 (43)               & 94.6 (13.1)               & 2.9 (2.6)               & 93.6 (9.2)                \\ \cline{1-2}
\multicolumn{1}{|c|}{\multirow{2}{*}{Different HP}}                                                              & Common grid   & 19.9 (54.7)             & 91.6 (17.8)               & 0.5 (0.4)               & 70.8 (24.3)               \\
\multicolumn{1}{|c|}{}                                                                                           & Uncommon grid & 14.5 (22.4)             & 89.1 (17.9)               & 2.5 (4.5)               & 81.1 (15.9)               \\ \cline{1-2}
\multicolumn{1}{|c|}{\multirow{2}{*}{\begin{tabular}[c]{@{}c@{}}Common HP on\\  different HP data\end{tabular}}} & Common grid   & 21.7 (36)               & 91 (19.8)                 & 1.5 (1.2)             & 91.1 (13)                 \\
\multicolumn{1}{|c|}{}                                                                                           & Uncommon grid & 18.1 (33)               & 92.5 (15.9)               & 3.2 (4.5)             & 93.4 (9.8)                \\ \hline
\end{tabular}
\end{center}
\end{table*}

\subsection{Running times comparisons}
\label{sec:simu_burden}

The counterpart of the more accurate and general results provided by \algo is a natural increase in running time.
\Cref{tab:simu_burden_compare} exhibits the raw and relative training times for GPFDA and \algo (prediction times are negligible and comparable in both cases), on data coming from the simulation scheme with varying values of $M$ on a \emph{Common grid} of $N = 30$ timestamps.
The algorithms were run under the \emph{3.6.1 R version}, on a laptop with a dual-core processor cadenced at 2.90GHz and an 8GB RAM.
The reported computing times are in seconds, and for small to moderate datasets ($N \simeq 10^3$, $M \simeq 10^4$ ) the procedures ran in few minutes to few hours. 
The difference between the two algorithms is due to GPFDA modelling $\mu_0$ as a deterministic function through B-splines smoothing, whereas \algo accounts for uncertainty. 
The ratio of computing times between the two methods tends to decrease as $M$ increases, and stabilises around $2$ for higher numbers of training individuals. 
This behaviour comes from the E step in \algo, which is incompressible and quite insensitive to the value of $M$.
Roughly speaking, one needs to pay twice the computing price of GPFDA for \algo to provide (significantly) more accurate predictions and uncertainty over $\mu_0$.
\Cref{tab:burden_diff_settings} provides running times of \algo according to its different settings, with $M=20$. 
Because the complexity is linear in $M$ in each case, the ratio in running times would remain roughly similar no matter the value of $M$. 
Prediction time appears negligible compared to training time, and generally takes less than one second to run. 
Besides, the \emph{Different HP} setting increases the running time since in this context $M$ maximisations (instead of one for \emph{Common HP}) are required at each EM iteration. 
In this case, the prediction also takes slightly longer because of the necessity to optimise hyper-parameters for the new individual.  
Although the nature of the grid of timestamps does not matter in itself, a key limitation lies in the dimension $N$ of the pooled set of timestamps, which tends to get bigger when individuals have different timestamps from one another.

\begin{table}[H]
\begin{center}
\caption{Average (standard deviation) training time (in seconds) for \algo and GPFDA on 100 runs for different numbers $M$ of individuals in the training dataset. The relative running time between {\algo} and GPFDA is provided on the line \emph{Ratio}.}
\label{tab:simu_burden_compare}
\begin{tabular}{c|cccc|}
\cline{1-5}
\multicolumn{1}{|c|}{$M =$ }  & 5            & 10           & 50           & 100          \\ \hline
\multicolumn{1}{|c|}{\algo}  & 5.2 (2.7)    & 7.6 (3.2)    & 24.2 (11.1)  & 42.8 (10)    \\
\multicolumn{1}{|c|}{GPFDA} & 1 (0.3)      & 2.1 (0.6)    & 10.7 (2.4)   & 23.1 (5.3)   \\ \hline
\multicolumn{1}{|c|}{Ratio} & \textbf{5.2} & \textbf{3.6} & \textbf{2.3} & \textbf{1.9} \\ \hline
\end{tabular}
\end{center}
\end{table}

\begin{table}[H]
\begin{center}
\caption{Average (standard deviation) training and prediction time (in seconds) on 100 runs for different settings of \algo.}
\label{tab:burden_diff_settings}
\begin{tabular}{cc|cc|}
\cline{3-4}
                                                    &               & Train    & Predict \\ \hline
\multicolumn{1}{|c|}{\multirow{2}{*}{Common HP}}    & Common grid   & 12.6 (3.5)  & 0.1 (0)    \\
\multicolumn{1}{|c|}{}                              & Uncommon grid & 16.5 (11.4) & 0.2 (0.1)  \\ \cline{1-2}
\multicolumn{1}{|c|}{\multirow{2}{*}{Different HP}} & Common grid   & 42.6 (20.5) & 0.6 (0.1)  \\
\multicolumn{1}{|c|}{}                              & Uncommon grid & 40.2 (17)   & 0.6 (0.1)  \\ \hline
\end{tabular}
\end{center}
\end{table}

\subsection{Application of \algo on swimmers' progression curves}
\label{sec:simu_real_data}

\paragraph{Data and problematic}
~\par

We consider the problem of performance prediction in competition for french swimmers. 
The French Swimming Federation provided us with an anonymised dataset, compiling the age and results of its members between 2000 and 2016. For each competitor, the race times are registered for competitions of 100m freestyle (50m swimming-pool). 
The database contains results from 1731 women and 7876 men, each of them compiling an average of 22.2 data points (min = 15, max = 61) and 12 data points (min = 5, max = 57), respectively.
In the following, age of the $i$-th swimmer is considered as the input variable (timestamp $t$) and the performance (in seconds) on a 100m freestyle as the output ($y_i(t)$).
For reasons of confidentiality and property, the raw dataset cannot be published.
The analysis focuses on the youth period, from 10 to 20 years, where the progression is the most noticeable. 
In order to get relevant time series, we retained only individuals having a sufficient number of data points ($N_i \geq 5$) on the considered time period. 
For a young swimmer, observed during its first years of competition, we aim at modelling its progression curve and make predictions on its future performance in the subsequent years. 
Since we consider a decision-making problem involving irregular time series, the GP probabilistic framework is a natural choice to work on.
Thereby, assuming that each swimmer in the database is a realisation $y_i$ defined as previously, we expect \algo to provide multi-task predictions for a new young swimmer, that will benefit from information of other swimmers already observed at older ages.
To study such modelling, and validate its efficiency in practice, we split the individuals into training and testing datasets with respective sizes: 

\begin{itemize}
	\item $M_{\operatorname{train}}^F = 1039$, for the female training set,
	\item $M_{\operatorname{test}}^F = 692$, for the female testing set,
	\item $M_{\operatorname{train}}^M = 4726$, for the male training set,
	\item $M_{\operatorname{test}}^M = 3150$, for the male testing set.
\end{itemize}

\noindent Inference on the hyper-parameters is performed thanks to the training dataset in both cases.
Considering the different timestamps and the relative monotony of the progression curves, the settings \emph{Uncommon grid}/\emph{Common HP} has been used for \algo. 
The overall training lasted around 2 hours with the same hardware configuration as for simulations. 
To compute MSE and the $CI_{95}$ coverage, the data points of each individual in the testing set has been split into \emph{observed} and \emph{testing} timestamps.
Since each individual has a different number of data points, the first 80\% of timestamps are taken as \emph{observed}, while the remaining 20\% are considered as \emph{testing} timestamps. 
{\algo}'s predictions are compared with the true values of $y_i$ at testing timestamps.
As previously, both GP and \algo have been initialised with a constant 0 mean function.
Initial values for hyper-parameters are also similar for all $i$, $\theta_0^{\operatorname{ini}} = \theta_i^{\operatorname{ini}} = (\exp(1), \exp(1))$ and $\sigma_i^{\operatorname{ini}} = 0.4$. 
Those values are the default in \algo and remain adequate in the context of these datasets.

\paragraph{Results and interpretation}

The overall performance and comparison are summarised in \Cref{tab:real_data}.

\begin{table}[H]
\caption{Average MSE (standard deviation) and average $CIC_{95}$ (standard deviation) for prediction on french swimmer testing datasets.}
\label{tab:real_data}
\begin{center}
\begin{tabular}{cc|cc|}
\cline{3-4}
\multicolumn{1}{l}{}                         &      & MSE                 & $CIC_{95}$          \\ \hline
\multicolumn{1}{|c|}{\multirow{2}{*}{Women}} & \algo & \textbf{3.8 (10.3)} & \textbf{95.3 (15.9)} \\
\multicolumn{1}{|c|}{}                       & GP   & 25.3 (97.6)         & 72.7 (37.1)          \\ \cline{1-2}
\multicolumn{1}{|c|}{\multirow{2}{*}{Men}}   & \algo & \textbf{3.7 (5.3)}  & \textbf{93.9 (15.3)} \\
\multicolumn{1}{|c|}{}                       & GP   & 22.1 (94.3)         & 78.2 (30.4)          \\ \hline
\end{tabular}
\end{center}
\end{table}

\begin{figure}[h]
    \begin{center}
       \includegraphics[width= \textwidth]{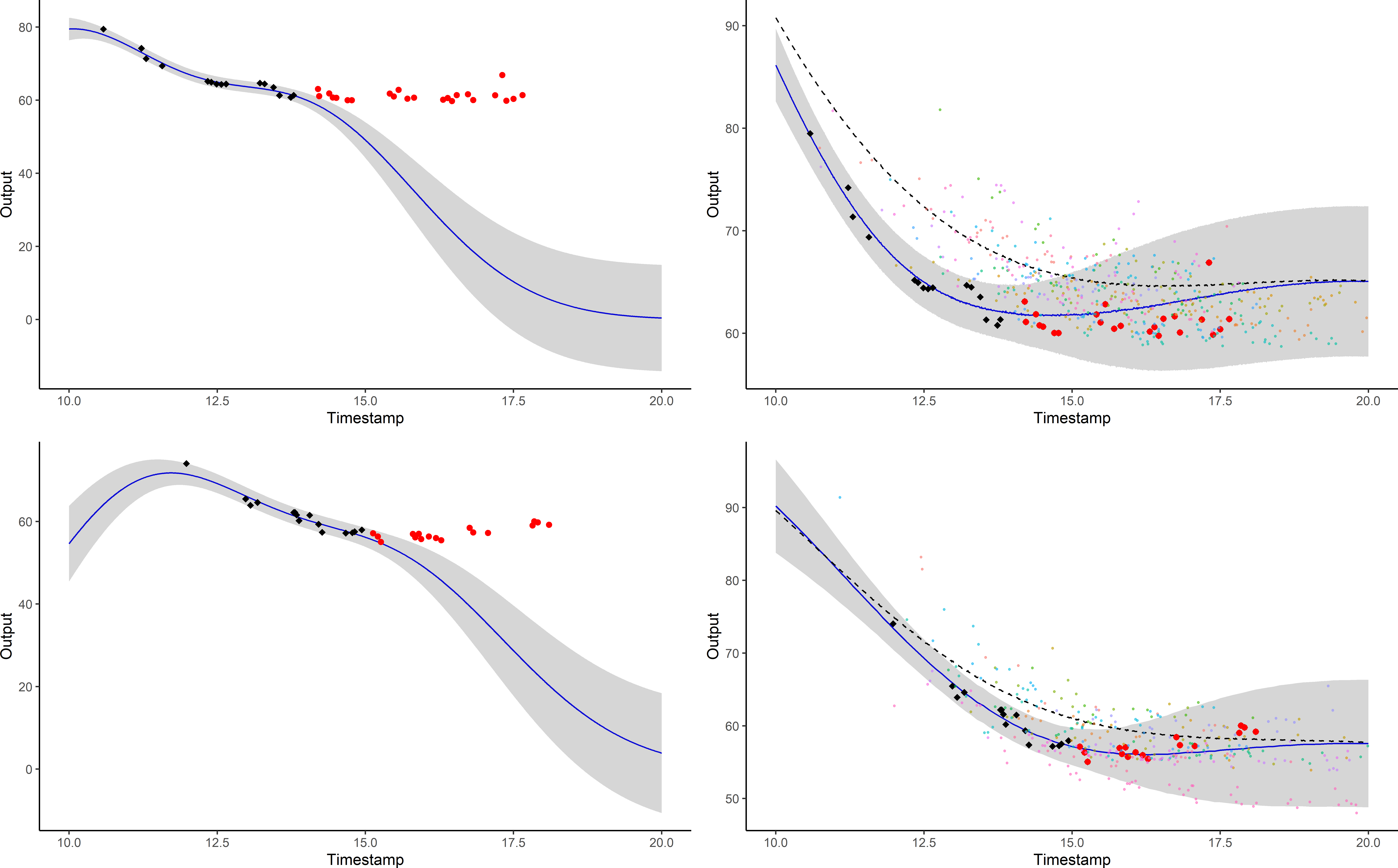}
       \caption{Prediction curves (blue) for a testing individual with associated 95\% credible intervals (grey) for GP regression (left) and \algo (right), for both women (top) and men (bottom). The dashed lines represent the mean functions $\hat{m}_0$, from the hyper-posterior $p(\mu_0 \vert \yii)$. Observed data points are in black, and testing data points are in red. The colourful backward points are observations from the training datasets, each colour corresponding to a different individual.}
		\label{fig:real_data}  
    \end{center}
\end{figure}

We observe that \algo still provides excellent results in this context, and naturally outperforms predictions provided by a standard GP regression. 
As the progression curves present relatively monotonic variations and thus avoid pathological behaviours that could occur with synthetic data, the MSE in prediction remains very low. 
The $CI_{95}$ coverage sticks close to the 95\% expected value for \algo, indicating an adequate quantification of uncertainty. 
To illustrate these results, an example is displayed on \Cref{fig:real_data} for both men and women. 
For a randomly chosen testing individual, we plot its predicted progression curve (in blue), where its first 15 data points are used as observations (in black), while the remaining true data points (in red) are displayed for comparison purpose. 
As previously observed in the simulation study, the simple GP quickly drifts to the prior 0 mean, as soon as data lack. 
However, for both men and women, the \algo predictions remain close to the true data, which also lie within the 95\% credible interval. 
Even for long term forecast, where the mean prediction curve tends to overlap the mean process (dashed line), the true data remain in our range of uncertainty, as the credible interval widens far from observations.
For clarity, we displayed only a few individuals from the training dataset (colourful points) in the background.
The mean process (dashed line) seems to represent the main trend of progression among swimmers correctly, even though we cannot numerically compare $\mu_0$ to any real-life analogous quantity.
From a more sport-related perspective, we can note that both genders present similar patterns of progression.
However, while performances are roughly similar in mean trend before the age of 14, they start to differentiate afterwards and then converge to average times with approximatively a 5 seconds gap. 
Interestingly, the difference between men and women in terms of world records in swimming competitions for the 100m freestyle is currently 4.8 seconds (46.91 versus 51.71).
These results, obtained under reasonable hypotheses on several hundreds of swimmers, seem to indicate that \algo would give quite reliable predictions for a new young swimmer. 
Furthermore, the uncertainty provided through the predictive posterior distribution offers an adequate degree of caution in a decision-making process.

\section{Discussion}
\label{sec:conclusion}

We have introduced a unified multi-task framework integrating a mean Gaussian process prior in the context of GP regression. 
While we believe that this process is an interesting object in itself, it also allows individuals to borrow information from each other and provides more accurate predictions, even far from data points. 
Furthermore, our method accounts for uncertainty in the mean process and remains applicable no matter eventual irregular timestamps in the grid of observations.
The proposed algorithm, \algo , also presents a reduced computational complexity compared to previous multi-task GPs frameworks.
Both on simulated and real-life datasets, we exhibited the efficiency of such an approach and studied some of its properties and possible settings. 
\algo outperforms the alternatives in estimation of the mean process as well as in prediction, and leads to a reliable quantification of uncertainty.
We also displayed evidence of its predictive efficiency for real-life problems and provided some insights on practical interpretations about the mean process. 
\newline

Despite the extensive literature on these aspects of GPs, our model does not yet include sparse approximations. 
While these aspects remain beyond the scope of the present paper, we might aim at adapting existing approaches \citep{SnelsonSparseGaussianProcesses2006,Quinonero-CandelaApproximationMethodsGaussian2007,TitsiasVariationalLearningInducing2009a} in our model to widen its applicability. 
Another possible avenue is an adaptation to the classification context, which is presented in \citet[][Chapter 3]{RasmussenGaussianprocessesmachine2006a}. 
Besides, this work leaves the door open to improvement as we tackled here the problem of unidimensional regression:
enabling either multidimensional or mixed type of inputs as in \cite{ShiGaussianProcessRegression2011} would be of interest. 
To conclude, the hypothesis of a unique underlying mean process might be considered too restrictive for some datasets, and enabling cluster-specific mean processes would be a relevant extension.

\section{Proofs}
\label{sec:proofs}

Note that the proof of \Cref{prop:E_step}  is a particular case of the proof below, where $\taub = \Ut$  exactly (where $\taub$ is the set of timestamps the hyper-posterior is to be computed on). 
Moreover, in order to keep an analytical expression for  $\mu_0$'s
hyper-posterior distribution, we discard the  superfluous information
contained in $\yii$ at timestamps  on which the hyper-posterior is not to be computed. 
Hence, the proof below states that the remaining data points are observed on subsets $\acc{\taub_i}_i$  of $\taub$.

\subsection{Proof of \Cref{prop:post_mu}}
\label{proof:E_step}

Let $\taub$  be a finite  vector of timestamps,  and $\acc{\taub_i}_i$ such as  $\forall i  \in \I ,   \ \  \taub_i \subset  \taub$. 
We define convenient notation:
 
\begin{itemize}
	\item $\mutau = \mu_0(\taub)$,
	\item $\mtau = m_0(\taub)$,
	\item $\mutaui = \mu_0(\taub_i), \ \forall i \in \I$,
	\item $\ytaui = y_i(\taub_i), \ \forall i \in \I$,
	\item $\boldsymbol{\Psi}_i = \psii(\taub_i, \taub_i), \forall i \in \I$,
	\item $\mathbf{K} = k_{\theta_0}(\taub, \taub)$.
\end{itemize}

Moreover, for a covariance matrix $C$, and $u, v \in \taub$, we note $\croch{C}_{uv}^{-1}$ the element of the inverse matrix at row associated with timestamp $u$, and column associated with timestamp $v$. 
We also  ignore the  conditionings over $\hat{\Theta}$, $\taub_i$ and
$\taub$ to maintain simple expressions.
By construction of the models, we have:
\begin{align*}
	p(\mutau \vert \acc{\ytaui}_i) 
	& \propto \displaystyle  p(\acc{\ytaui}_i\vert \mutau) p( \mutau)    \\ 
	& \propto \left\{\displaystyle \prod_{i = 1}^{M} p(\ytaui \vert \mutaui)\right\} p(\mutau)  \\ 
	& \propto  \left\{\displaystyle \prod_{i = 1}^{M} \mathcal{N} \paren{\ytaui; \mutaui, \boldsymbol{\Psi}_i)}\right\} \mathcal{N} \paren{ \mutau; \mtau, \mathbf{K}}.   
\end{align*}

The term  $\mathcal{L}_1 = - (1/2)\log  p(\mutau \vert \acc{\ytaui}_i)
$ associated with the hyper-posterior remains quadratic and we may find
the corresponding Gaussian parameters by identification:

\begin{align*}
	\mathcal{L}_1
	&= \sumi \acc{\paren{\ytaui - \mutaui}^{\intercal} \boldsymbol{\Psi}_i^{-1} \paren{\ytaui - \mutaui} + C_i } + \paren{\mutau - \mtau}^{\intercal} \mathbf{K}^{-1} \paren{\mutau - \mtau} + C_0 \\
	&= {\mutau}^{\intercal} \mathbf{K}^{-1} \mutau + \sumi {\mutaui}^{\intercal} \boldsymbol{\Psi}_i^{-1} \mutaui - 2 \paren{ {\mutau}^{\intercal} \mathbf{K}^{-1} \mtau + \sumi {\mutaui}^{\intercal} \boldsymbol{\Psi}_i^{-1} \ytaui } + C \\
	&= \sumu \sumv \mu_0(u) \croch{\mathbf{K}}_{uv}^{-1} \mu_0(v)  + \sumi \sum\limits_{u \in \taub_i} \sum\limits_{v \in \taub_i} \mu_0(u) \croch{\boldsymbol{\Psi}_i}_{uv}^{-1} \mu_0(v)  \\ 	
	& \ \ \ - 2 \sumu \sumv  \mu_0(u) \croch{\mathbf{K}}_{uv}^{-1} m_0(v) - 2 \sumi \sum\limits_{u \in \taub_i} \sum\limits_{v \in \taub_i} \mu_0(u)  \croch{\boldsymbol{\Psi}_i}_{uv}^{-1} y_i(v) + C,
\end{align*}

\noindent where we entirely decomposed the vector-matrix products.
We factorise the expression according to the common timestamps between $\tau_i$ and $\tau$. 
Since for all $ i, \taub_i \subset \taub$, let us introduce a dummy indicator function $\mathds{1}_{\taub_i} = \mathds{1}_{\{u,v \in \taub_i\}}$ to write: 

\begin{align*}
	\sumi \sum\limits_{u \in \taub_i} \sum\limits_{v \in \taub_i} A(u,v) 
	&= \sumi \sumu \sumv \mathds{1}_{\taub_i} A(u,v) \\
	&= \sumu \sumv \sumi \mathds{1}_{\taub_i} A(u,v),
\end{align*}

\noindent subsequently, we can gather the sums such as:

\begin{align*}
	\mathcal{L}_1
	&= \sumu \sumv \Bigg( \mu_0(u) \croch{\mathbf{K}}_{uv}^{-1} \mu_0(v) + \sumi \mathds{1}_{\taub_i} \mu_0(u) \croch{\boldsymbol{\Psi}_i}_{uv}^{-1} \mu_0(v) \\ 	
	& \hspace{1.8cm} - 2  \mu_0(u) \croch{\mathbf{K}}_{uv}^{-1} m_0(v) - 2 \sumi \mathds{1}_{\taub_i} \mu_0(u)  \croch{\boldsymbol{\Psi}_i}_{uv}^{-1} y_i(v) \Bigg) + C \\
	&= \sumu \sumv \Bigg( \mu_0(u) \Big( \croch{\mathbf{K}}_{uv}^{-1} + \sumi \mathds{1}_{\taub_i}  \croch{\boldsymbol{\Psi}_i}_{uv}^{-1} \Big) \mu_0(v) \\ 	
	& \hspace{1.8cm} - 2  \mu_0(u) \Big( \croch{\mathbf{K}}_{uv}^{-1} m_0(v) + \sumi \mathds{1}_{\taub_i} \croch{\boldsymbol{\Psi}_i}_{uv}^{-1} y_i(v) \Big) \Bigg) + C \\
	&= {\mutau}^{\intercal} \Big( \mathbf{K}^{-1} + \sumi \tilde{\boldsymbol{\Psi}}_i^{-1} \Big) \mutau  -2 {\mutau}^{\intercal} \Big( \mathbf{K}^{-1} \mtau + \sumi \tilde{\boldsymbol{\Psi}}_i^{-1} \tilde{\mathbf{y}}_i^{\taub} \Big)+ C,
\end{align*}

\noindent where the $\yi$ and $\boldsymbol{\Psi}_i$ are completed by zeros: 

\begin{itemize}
	\item $\tilde{\mathbf{y}}_i^{\taub} =  \mathds{1}_{\taub_i} y_i( \taub  ) $, 
	\item $\croch{ \tilde{\boldsymbol{\Psi}}_i}_{uv}^{-1} = \mathds{1}_{\taub_i}  \croch{\boldsymbol{\Psi}_i}_{uv}^{-1}, \ \forall u,v \in \taub$.
\end{itemize}

By identification of the quadratic form, we reach: 

\begin{equation*}
	p(\mutau \vert \acc{\ytaui}_i) = \mathcal{N} \paren{ \mutau ; \mhat(\taub), \Khat },
\end{equation*}

\noindent with,

\begin{itemize}
	\item $\Khat = \paren{ \mathbf{K}^{-1} + \sumi \tilde{\boldsymbol{\Psi}}_i^{-1} }^{-1}$,
	\item $\mhat(\taub) = \Khat \paren{ \mathbf{K}^{-1} \mtau + \sumi \tilde{\boldsymbol{\Psi}}_i^{-1} \tilde{\mathbf{y}}_i^{\taub} } $.
\end{itemize}
\qed

\subsection{Proof of \Cref{prop:M_step_diff} and \Cref{prop:M_step_common}}
\label{proof:Proof_M_step}

Since the central part of the proofs is similar for both propositions, we detail the calculus by denoting $\Theta = \{ \theta_0, \allowbreak \thetaii, \sigmaii \}$ for generality, and dissociating the two cases only when necessary. 
Before considering the maximisation, we notice that the joint density can be developed as:

\begin{align*}
	p(\yii, \mu_0(\Ut) \vert \Theta)
 	 &= p(\yii \vert \mu_0(\Ut) , \Theta) \ p(\mu_0(\Ut) \vert \Theta) \\
	 &= \prod_{i=1}^{M}\left\{ p(\yi \vert \mu_0(\Ut) , \theta_i, \sigma_i^2)\right\} \ p(\mu_0(\Ut) \vert \theta_0) \\
	 &= \prod_{i=1}^{M} \left\{\mathcal{N}(\yi; \mu_0(\Ut) , \Psii) \right\} \mathcal{N}(\mu_0(\Ut); m_0(\Ut), \mathbf{K}_{\theta_0}^{\Ut}).
\end{align*}

The expectation  is taken  over $p(\mu_0(\Ut)  \vert \yii)$  though we
write it $\mathbb{E}$ for simplicity. We have:
\begin{align*}
	f(\Theta)
	&=\mathbb{E} \croch{ \log p(\yii, \mu_0(\Ut) \vert \Theta) } \\
	&= - \dfrac{1}{2} \mathbb{E}\Bigg[ (\mu_0(\Ut) - m_0(\Ut))^{\intercal} {\mathbf{K}_{\theta_0}^{\Ut}}^{-1} (\mu_0(\Ut) - m_0(\Ut)) -  \log \abs{{\mathbf{K}_{\theta_0}^{\Ut}}^{-1}} \\
	& \hspace{1.5cm} +  \sumi (\yi - \mu_0(\ti))^{\intercal} {\Psii^{\ti}}^{-1} (\yi - \mu_0(\ti)) -  \log \abs{{\Psii^{\ti}}^{-1}} \Bigg] + C_1.
\end{align*}

\begin{lemma}
\label{lem:1}
	Let $X \in \mathbb{R}^N$ be a random Gaussian vector $X \sim \mathcal{N} \paren{ m , \mathbf{K} }$, $b \in \mathbb{R}^N$, and $\mathbf{S}$, a $N \times N$ covariance matrix. Then:
	
	\begin{align*}
		E &=\mathbb{E}_{X } \croch{ (X - b)^{\intercal} \mathbf{S}^{-1} (X - b)} \\
		&= (m - b)^{\intercal} \mathbf{S}^{-1} (m - b) + \operatorname{Tr}(\mathbf{K} \mathbf{S}^{-1}).
	\end{align*}
\end{lemma}

\begin{proof}[\Cref{lem:1}]

	\begin{align*}
		E &=              \mathbb{E}_{X              }\croch{\operatorname{Tr}
           (\mathbf{S}^{-1}(X-b)(X-b)^{\intercal})}\\
                  &= \operatorname{Tr}(\mathbf{S}^{-1}\mathbb{V}_{X}(X-b))+\operatorname{Tr}(\mathbf{S}^{-1} (m-b)(m-b)^{\intercal})\\
		&=  (m - b)^{\intercal}	\mathbf{S}^{-1}(m - b) + \operatorname{Tr} \paren{ \mathbf{K} \mathbf{S}^{-1} }.
	\end{align*}
\end{proof}

As we note that $X$ and $b$ play symmetrical roles in the calculus of the conditional expectation, we can apply the lemma regardless of the position of $\mu_0$ in the $M+1$ equalities involved.
Applying \Cref{lem:1} to our previous expression of $f(\Theta)$, we obtain:

\begin{align*}
	f(\Theta)
	&= - \dfrac{1}{2} \Bigg[ (\mhat(\Ut) - m_0(\Ut))^{\intercal} {\mathbf{K}_{\theta_0}^{\Ut}}^{-1} (\mhat(\Ut) - m_0(\Ut)) \\
	& \ \ \ +  \sumi (\yi - \mhat(\ti))^{\intercal} {\Psii^{\ti}}^{-1} (\yi - \mhat(\ti)) \\  
	& \ \ \ +  \operatorname{Tr} \paren{ \Khat^{\Ut} {\mathbf{K}_{\theta_0}^{\Ut}}^{-1}} +  \sumi \operatorname{Tr} \paren{ \Khat^{\ti} {\Psii^{\ti}}^{-1} } \\
	& \ \ \ - \log \abs{{\mathbf{K}_{\theta_0}^{\Ut}}^{-1}} - \sumi \log \abs{{\Psii^{\ti}}^{-1}} + C_1 \Bigg].
\end{align*}

We recall that, at the M step, $\mhat(\Ut)$ is a known constant, computed at the previous E step. 
Thus, we identify here the characteristic expression of several Gaussian log-likelihoods and associated correction trace terms. 
Moreover, each set of hyper-parameters is merely involved in independent terms of the whole function to maximise. 
Hence, the global maximisation problem can be separated into several maximisations of sub-functions according to the hyper-parameters getting optimised. 
Regardless to additional assumptions, the hyper-parameters $\theta_0$, controlling the covariance matrix of the mean process, appears in a function which is exactly a Gaussian log-likelihood, $\log\mathcal{N} \paren{ \mhat(\Ut), m_0(\Ut) , \mathbf{K}_{\theta_0}^{\Ut} }$, added to a corresponding trace term, $ - \dfrac{1}{2} \operatorname{Tr} \paren{ \Khat^{\Ut} {\mathbf{K}_{\theta_0}^{\Ut}}^{-1}}$. 
This function can be maximised independently from the other parameters, giving the first part of the results in \Cref{prop:M_step_diff} and \Cref{prop:M_step_common}. 
\newline

Although the idea is analogous for the remaining hyper-parameters, we have to discriminate here regarding the assumption on the model. 
If each individual is supposed to have its own set $\acc{\theta_i, \sigma_i}$, which thus can be optimised independently from the observations and hyper-parameters of other individuals, we identify a sum of $M$ Gaussian log-likelihoods, $\log \mathcal{N} \big( \yi , \mhat(\ti) , \Psii^{\ti} \big)$, and the corresponding trace terms, $- \dfrac{1}{2} \operatorname{Tr}( \Khat^{\Ut} {\Psii^{\ti}}^{-1} )$. 
This property results in $M$ independent maximisation problems on the corresponding functions, proving \Cref{prop:M_step_diff}.
Conversely, if we assume that all individuals in the model shares their hyper-parameters (i.e. $\acc{ \theta , \sigma^2 } = \acc{ \theta_i, \sigma_i^2 }, \forall i \in \I$),  we can no longer divide the problem into $M$ sub-maximisations, and the whole sum on all individual should be optimised thanks to observations from all individuals. This case corresponds to the second part of \Cref{prop:M_step_common}. 
\qed

\section*{Acknowledgements}
The authors warmly thank Andy Marc, Olivier Dupas, Richard Martinez and the French Swimming Federation for providing data and helping in the analysis of the results. 
Benjamin Guedj acknowledges partial support by the U.S. Army Research Laboratory and the U.S. Army Research Office, and by the U.K. Ministry of Defence and the U.K. Engineering and Physical Sciences Research Council (EPSRC) under grant number EP/R013616/1. 
Benjamin Guedj acknowledges partial support from the French National Agency for Research, grants ANR-18-CE40-0016-01 and ANR-18-CE23-0015-02.

%
\section*{Declarations}

\paragraph{Funding.}
The authors received no financial support for the research, authorship, and/or publication of this article.

\paragraph{Conflict of interest.}
The authors declare no conflict of interest.

\paragraph{Availability of data.}
The synthetic data and table of results are available at \url{https://github.com/ArthurLeroy/MAGMA/tree/master/Simulations}.
 
\paragraph{Code availability.}
The R code associated with the present work is available at \url{https://github.com/ArthurLeroy/MAGMA}.
The current version of the R package implementing an extended version of \algo is available at \url{https://github.com/ArthurLeroy/MagmaClustR}.

\bibliography{biblio}   

\end{document}